\documentclass[preprintnumbers,amsmath,amssymb]{revtex4}
\usepackage{graphicx}
\usepackage{bm}

\begin{document}

\title{Transformation of polar nematic phases in the presence of electric field}
\author{A. V. Emelyanenko$^{1}$}
\email{emel@polly.phys.msu.ru}
\author{V. Yu. Rudyak$^{1}$}
\author{F. Araoka$^{2}$}
\author{H. Nishikawa$^{2}$}
\author{K. Ishikawa$^{3}$}
\affiliation{$^1$Lomonosov Moscow State University, Moscow 119991,
Russia \\ $^2$RIKEN Center for Emergent Matter Science (CEMS), 2-1
Hirosawa Wako, Saitama 351-0198 Japan \\ $^3$Tokyo Institute of
Technology, 2-12-1 Ookayama, Meguro-ku Tokyo 152-8550 Japan}

\begin{abstract}
Only a few years have passed since discovery of polar nematics,
and now they are becoming the most actively studied liquid crystal
materials. Despite numerous breakthrough findings made recently, a
theoretical systematization is still lacking. In the present paper
we are making a step on the way of systematization. A powerful
technique that molecular-statistical physics is has been applied
to an assembly of polar molecules influenced by electric field.
Totally, the three polar nematic phases were found to be stable at
various conditions: the double-splay ferroelectric nematic
$N_F^{2D}$ (observed in the lower-temperature range in the absence
or at low electric field), the double-splay antiferroelectric
nematic $N_{AF}$ (observed at intermediate temperature in the
absence or at low electric field) and the single-splay
ferroelectric nematic $N_F^{1D}$ (observed at moderate electric
field at any temperature below transition into paraelectric
nematic $N$ and in the higher-temperature range (also below $N$)
at low electric field or without it. A paradoxal transition from
$N_F^{1D}$ to $N$ induced by application of higher electric field
has been found and explained. A transformation of the structure of
polar nematic phases at application of electric field has also
been investigated by Monte Carlo simulations and experimentally by
observation of POM images. In particular, it has been realized
that, at planar anchoring, $N_{AF}$ in the presence of moderate
out-of-plane electric field exhibits the twofold splay modulation:
antiferroelectric in the plane of the substrate and ferroelectric
in the plane normal to the substrate. Several additional
sub-transitions related to fitting confined geometry of the cell
by the structure of polar phases were detected.
\end{abstract}

\maketitle

\section{Introduction}

One of the main trends in modern science today is the development
of new materials, which can be effectively manipulated by electric
field for various humans needs, from displays to medicine. Liquid
crystals (LCs) fulfill many demands. It was noticed that LCs
possessing spontaneous polarization can be better candidates for
the novel applications: from fast energy-saving and compact
electronics to artificial muscles. However, the layered structures
of smectics (the only class of LCs known previously to posses the
spontaneous polarization) are poorly resistant to mechanical
stress. Over the past few years, several new classes of nematic
LCs (which are sustainable to mechanical stress) with unique properties originating from unique symmetry of
individual molecules have been discovered.

For the last decades, the formation of spontaneous polarization in
the nematic materials has been actively discussed
\cite{Mandle:2022a,Vanakaras:2003,Takezoe:2014,Takezoe:2017}.
Liquid crystals consisting of bent-core molecules were considered
as the main candidates, since they have a significantly (several
orders of magnitude) higher flexoelectric coefficient
\cite{Vita:2018}. Indeed, nanosized polar clusters in the nematic
phase were found for this kind of mesogens
\cite{Francescangeli:2010}. However, these materials do not
possess macroscopic polarization in the absence of external field.
Meanwhile, proper ferroelectricity was found in columnar phases
composed of umbrella-shaped mesogens
\cite{Gorecka:2004,Fitie:2010,Miyajima:2012} and in re-entrant
smectic phases
\cite{Novotna:2001,Catalano:2006,Bubnov:2008,Na:2008,Emelyanenko:2019}.

In 2017, the two scientific groups independently reported about
the existence of polar nematic phases in LCs composed of the
wedge-shaped molecules
\cite{Nishikawa:2017,Mandle:2017a,Mandle:2017b}. In
Ref.~\cite{Mertelj:2018,Mandle:2019} it was realized that the
polar nematic phases can demonstrate spontaneous splay and
flexoelectricity, while the existence of splay flexoelectricity in
nematic LCs was predicted earlier theoretically in
Ref.~\cite{Dhakal:2010}. It was also noticed in
Ref.~\cite{Mandle:2021a,Mandle:2021b,Cruickshank:2022,Li:2022,Mandle:2022b}
that minor changes in the molecular shape can sufficiently modify
the phase sequence.

In Ref.~\cite{Nishikawa:2017} it was demonstrated that DIO
material possesses at least three nematic phases: conventional
paraelectric nematic phase ($N$ or $M1$) at higher temperature,
ferroelectric nematic phase ($N_F$ or $MP$) at lower temperature
and some intermediate phase ($N_X$ or $M2$) in between them. In
Refs.~\cite{Mandle:2017a,Mandle:2017b} and later in
Ref.~\cite{Brown:2021} it was confirmed that RM-734 material
demonstrates $N_F$, but does not demonstrate $N_X$. The
anomalously high dielectric permittivity and dielectric anisotropy
were found in the $N_F$ phase in both DIO and RM-734 materials
\cite{Li:2021}. The value of spontaneous polarization in $N_F$ is
comparable with that for the solid-state ferroelectrics. At
present, many other polar nematic (including chiral and biaxial)
phases were found in different materials and mixtures
\cite{Sebastian:2022,Ortega:2022,Nishikawa:2021,Zhao:2021}. Our
theoretical studies presented in Ref.~\cite{Emelyanenko:2022}
suggest that the intermediate $M2$ (or $N_X$) phase observed in
DIO can be the antiferroelectric double splay nematic phase (in
correspondence with definition and in complete agreement with
Refs.~\cite{Rosseto:2020,Selinger:2022}) forming the periodical
2D-splay domains of several micrometers size. The same conclusions
follow from dielectric measurements and POM observations in
Ref.~\cite{Nishikawa:2017,Sebastian:2020} and measurements of
spontaneous polarization and PFM observations in
Ref.~\cite{Brown:2021}.

Experimentally, it is becoming more and more evident
\cite{Sebastian:2023} that ferroelectric nematic phase ($MP$ or $N_F$)
also possesses the splay domains. For the uniformity of
description, here and below we are going to use the $N_F^{1D}$ and
$N_F^{2D}$ notations by combining ``$N_F$'' with ``single-splay''
or ``double-splay'' definitions introduced in
Refs.~\cite{Rosseto:2020,Selinger:2022}). To be consistent
furthermore, we are going to use the $N_{AF}$ notation for $M2$
($N_X$), which is antiferroelectric.

There are many expectations about applications of nematic
ferroelectrics (NFs). Generally, their behavior at application of
electric field is not trivial, in particular, in combination of
electric field and surface-related effects \cite{Mathe:2023}. NFs
are the good candidates in nonlinear optics \cite{Folcia:2022}.
Interesting effects related to the motion of ferroelectric nematic
droplets in isotropic melts are considered in
Ref.~\cite{Perera:2023}, and the light-induced branched structures
of NF droplets on the surfaces are observed in
Ref.~\cite{Sterle:2023}. Various polarization topologies in
confined NFs were discussed in Ref.~\cite{Yang:2022}.

\begin{figure}[h!]
\includegraphics[width=0.4\linewidth,clip]{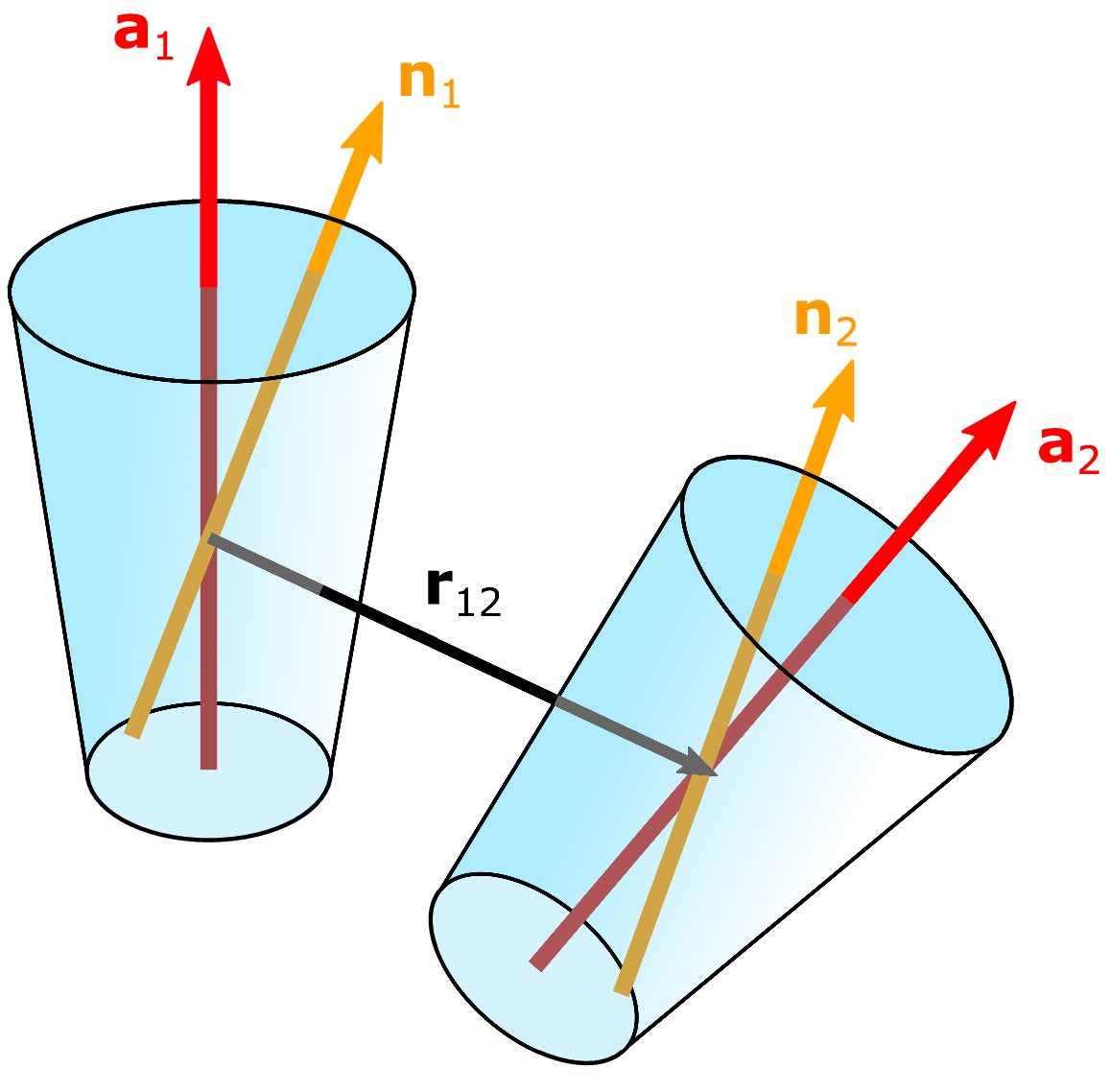}
\caption{\label{fig:epsart} A pair of interacting polar molecules.
Adapted from Ref.~\cite{Emelyanenko:2022}.}
\end{figure}

The discussion on how many polar nematic phases can exist, which of them are
splayed and which are uniform, which of them are proper and which
are improper, is still actual. Recently in Ref.~\cite{Saha:2022}
it was reported about the existence of three kinds of NF phases.
In the present paper we justify the existence of three
(macroscopically uniaxial and achiral) polar nematic phases. In
particular, we expect that all three polar phases can be observed
in DIO material. We are going to present theoretical explanations,
why all three polar phases are splayed and improper ferroelectric.
At the same time, since the splay domain size can achieve several
micrometers, in some temperature ranges, the splay director
deformation can be suppressed by the surfaces if the cell
thickness is lower than the domain size.

The paper is organized as follows. In Sec. II the structures of
polar nematic phases merged from theory, computer simulations and
experiment will be outlined and systemized. In Sec. III the
transformations of polar nematic phases induced by variation of
temperature and electric field will be investigated. In Sec. IV
the theoretical approaches used for analysis of the structures of
polar nematic phases will presented. Finally, in Sec. V the
conclusions will be made.

\section{The structures of polar nematic phases merged from theory, computer simulations and experiment}
\subsection{Generalization of elastic free energy for the presence of flexoelectric and induced polarizations}
It is known that flexoelectric effect is crucial for the formation
of various polar nematic phases. We are considering the polar
molecules similarly to that presented in
Ref.~\cite{Emelyanenko:2022}. Technically, the flexoelectric term
in the free energy can be obtained from consideration of specific
symmetry of the pair molecular potential. In particular, the
effective pair molecular interaction potential $U_{12}^{ef}({\bf
a}_1,{\bf a}_2,{\bf r}_{12})$ can be approximated by spherical
invariants $T_{\ell\,L\,\lambda}({\bf a}_1,{\bf u}_{12},{\bf
a}_2)$, where ${\bf a}_1$ and ${\bf a}_2$ are the principal axes
of molecules $1$ and $2$ located at points ${\bf r}_1$ and ${\bf
r}_2$, respectively, and ${\bf u}_{12}\equiv{\bf r}_{12}/|r_{12}|$
is the unit intermolecular vector, ${\bf r}_{12}\equiv{\bf
r}_2-{\bf r}_1$ (Fig.~1):
\begin{equation}
U^{ef}_{12}({\bf a}_1,{\bf a}_2,{\bf r}_{12})=
-\sum_{\ell,L,\lambda} J_{\ell L \lambda}(r_{12}) T_{\ell L
\lambda}({\bf a}_1,{\bf u}_{12},{\bf a}_2) \quad. \quad
\end{equation}
Introducing the polar $P({\bf r})$ and non-polar $S({\bf r})$
orientational order parameters
\begin{eqnarray}
P({\bf r})\equiv\int f[({\bf a}\cdot{\bf n}),{\bf r}] P_1({\bf
a}\cdot{\bf n}) d^2 {\bf a} \quad, \quad  S({\bf
r})\equiv\int f[({\bf a}\cdot{\bf n}),{\bf r}] P_2({\bf
a}\cdot{\bf n}) d^2 {\bf a} \quad, \quad
\end{eqnarray}
\begin{figure*}[tbh!]
\includegraphics[width=0.8\linewidth,clip]{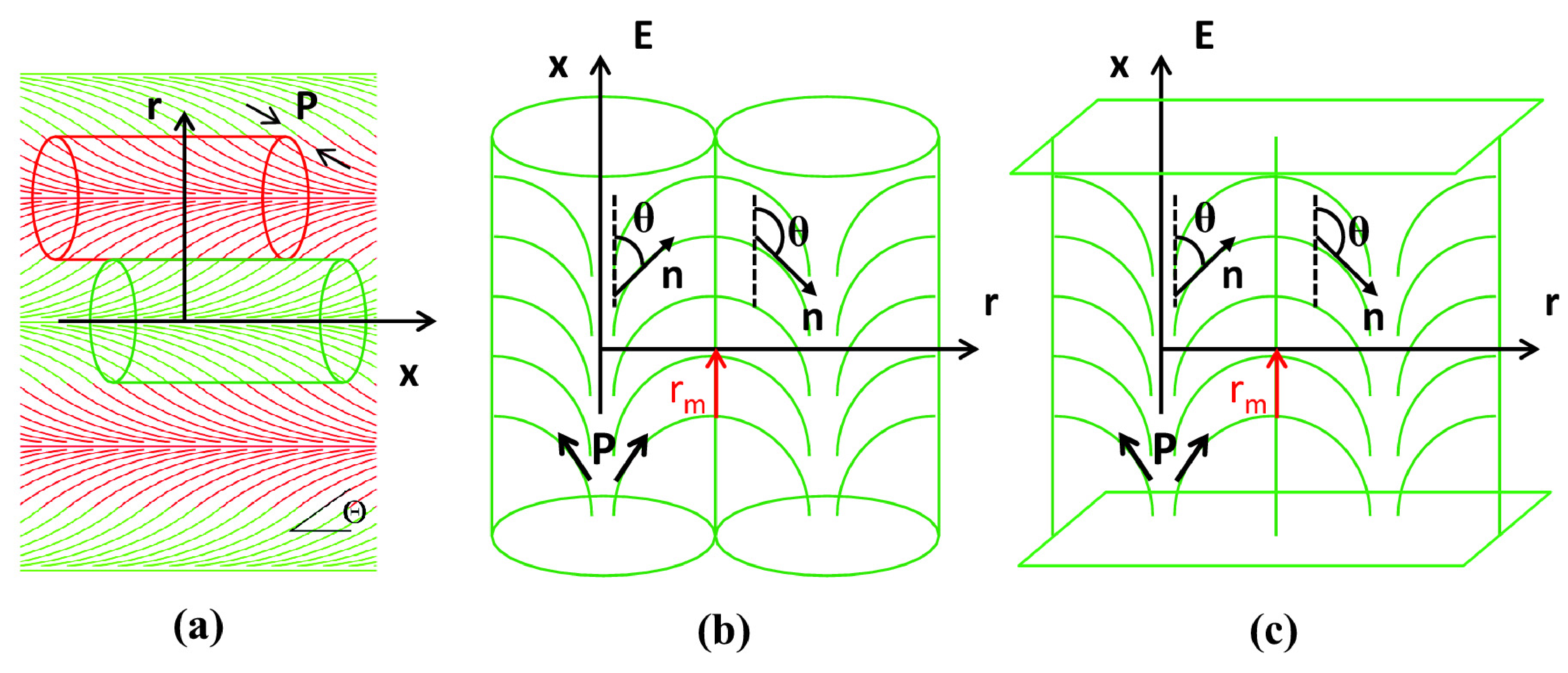}
\caption{\label{fig:epsart} Director distribution in $N_{AF}$ (a),
$N_F^{2D}$ (b) and $N_F^{1D}$ (c). Green color corresponds to the
positive splay and polarization, red color corresponds to the
negative splay and polarization; ${\bf x}$-axis is either along
the domain symmetry axis in (a) and (b) or along the symmetry
plane in (c); ${\bf r}$-axis is perpendicular to ${\bf x}$-axis;
$\theta$ is the angle between the local director ${\bf n}$ and
${\bf x}$-axis; ${\bf P}$ is the local polarization.}
\end{figure*}
where $f[({\bf a}\cdot{\bf n}),{\bf r}]$ is the orientational
distribution function for molecules having principal axes ${\bf
a}$ with respect to director ${\bf n}$ at point ${\bf r}$, and
using the gradient expansion of the director
\cite{Emelyanenko:2021,Emelyanenko:2015}, we obtain the
flexoelectric term as the average of $T_{110}({\bf a}_1,{\bf
u}_{12},{\bf a}_2)$ and $T_{011}({\bf a}_1,{\bf u}_{12},{\bf
a}_2)$ polar invariants \cite{Emelyanenko:2022}:
\begin{eqnarray}
\langle J_{110}(r_{12})T_{110}({\bf a}_1,{\bf u}_{12},{\bf a}_2)
 +J_{011}(r_{12})T_{011}({\bf a}_1,{\bf u}_{12},{\bf a}_2)\rangle\Longrightarrow \lambda P ({\mbox {\boldmath $\nabla$}}\cdot{\bf n}) \quad,\quad
\end{eqnarray}
where $\lambda$ is proportional to flexoelectric coefficient. The elastic free-energy density can be generalized by inclusion of
flexoelectric splay term and the term related to the presence of external electric
field:
\begin{eqnarray}
\frac{\partial F_{\bf n}}{\partial V}=\frac{1}{2}K_{11}\,\{{\bf
n}\,({\mbox {\boldmath $\nabla$}}\cdot{\bf n})- \lambda {\bf
P}\}^2 +\frac{1}{2}\,K_{22}({\bf
n}\cdot[{\mbox {\boldmath $\nabla$}}\times{\bf n}])^2
+\frac{1}{2}K_{33}[{\bf n}\times[{\mbox {\boldmath
$\nabla$}}\times{\bf n}]]^2 -\varepsilon_a ({\bf
E}\cdot{\bf P})
\quad,\quad
\end{eqnarray}
where ${\bf P}({\bf r})$ is the vector having absolute value
$P({\bf r})$ [see definition in Eq.~(2)] and (at positive $P$)
parallel to particular direction (one of the two opposite
directions) of pseudovector ${\bf n}({\bf r})$, $K_{11}$, $K_{22}$
and $K_{33}$ are the splay, twist and bend elastic constants,
respectively, $K_{11}\lambda$ is the flexoelectric constant
[Eq.~(3)], and $\varepsilon_a$ is the dielectric anisotropy of the
material.   At positive $\lambda$, polarization ${\bf P}$ is
parallel to director ${\bf n}$ at positive splay $({\mbox
{\boldmath $\nabla$}}\cdot{\bf n})$, and is anti-parallel to ${\bf
n}$ at negative splay. Here we should note that Eq.~(4) is only a part of the free-energy density, which explicitly depends on director ${\bf n}$, but it does not contain all the terms depending on the polarization value $P$. The total free-energy density will be considered in Sec. IV.

\subsection{Equilibrium structures of polar nematic phases}
To obtain the equilibrium structures of polar nematic material at
various conditions, we should minimize the total free energy independently with
respect to director ${\bf n}(\bf r)$ and orientational distribution function $f[({\bf a}\cdot{\bf n}),{\bf r}]$. The whole director-dependent part of the free-energy density is presented
in Eq.~(4),
while the total free-energy density depending explicitly on the orientational distribution function $f[({\bf a}\cdot{\bf n}),{\bf r}]$ will be considered in the framework of molecular statistical theory in Sec.~IV~A.
One notes, however, that Eq.~(4) contains also polarization ${\bf P}({\bf r})$, which is determined by function $f[({\bf a}\cdot{\bf n}),{\bf r}]$ in correspondence with Eq.~(2), and therefore the director and the orientational distribution function are correlated. This correlation will be considered in the framework of perturbation theory in
Sec.~IV~B.
\begin{figure}[h!]
\includegraphics[width=0.45\linewidth,clip]{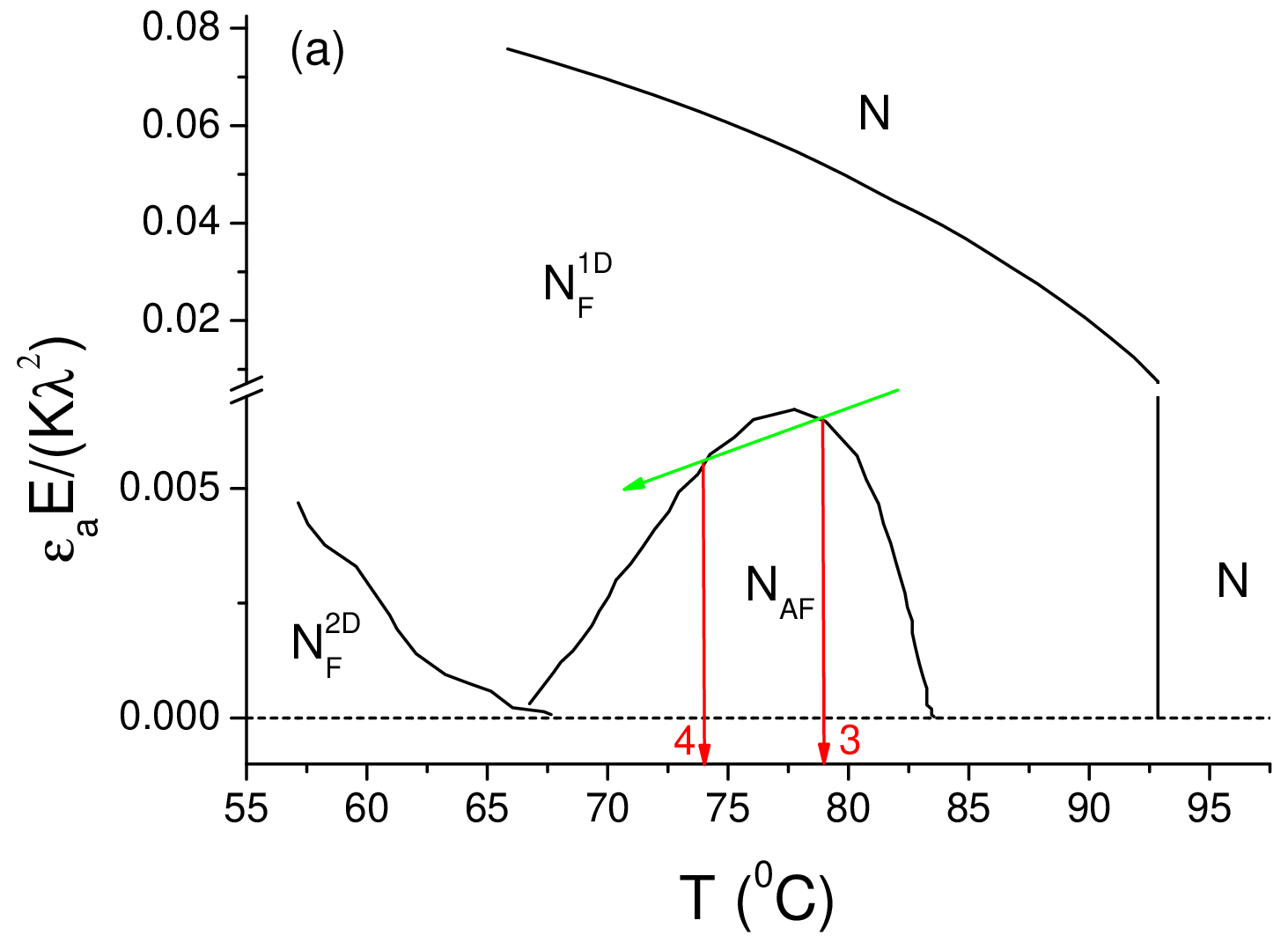}
\includegraphics[width=0.45\linewidth,clip]{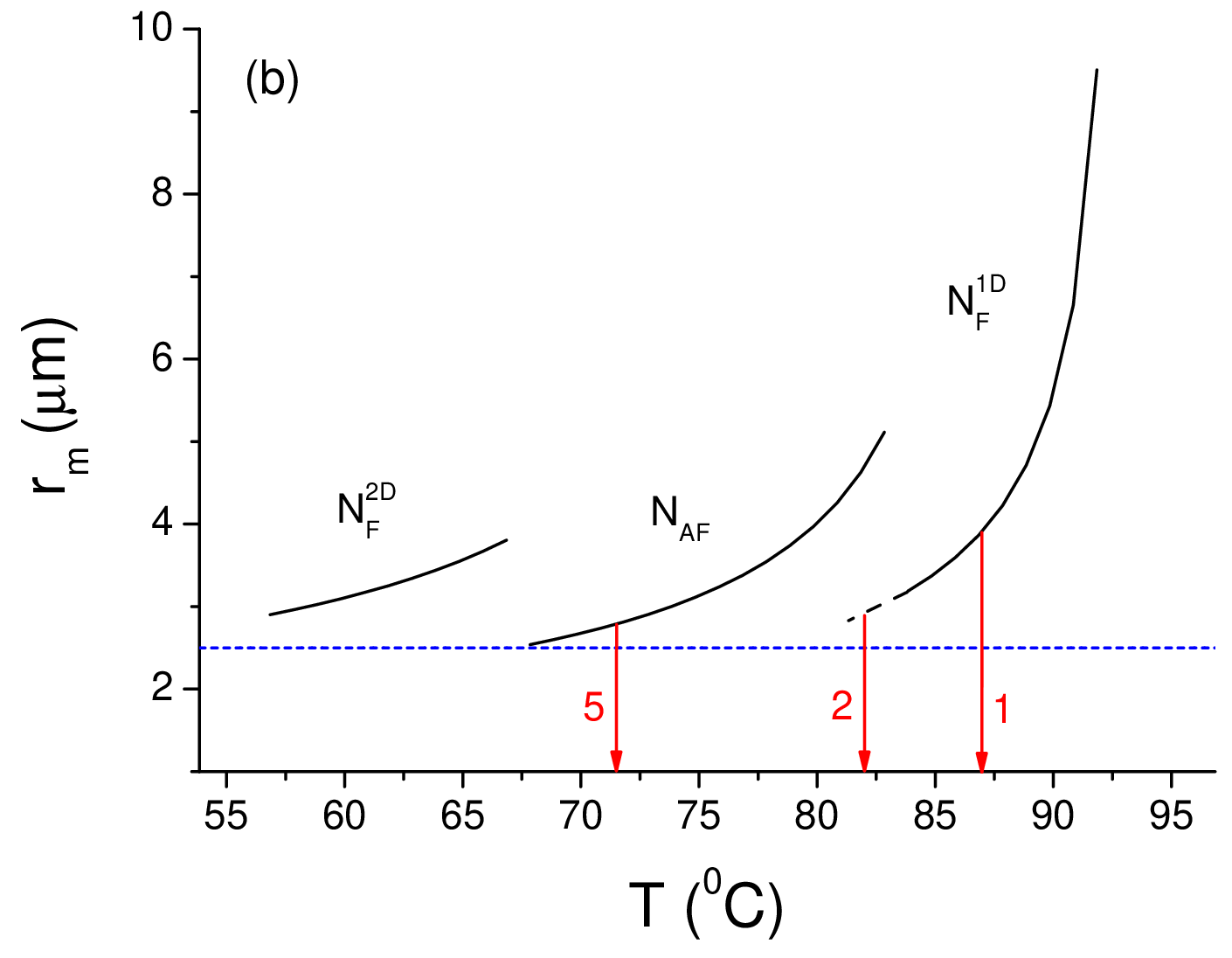}
\includegraphics[width=0.45\linewidth,clip]{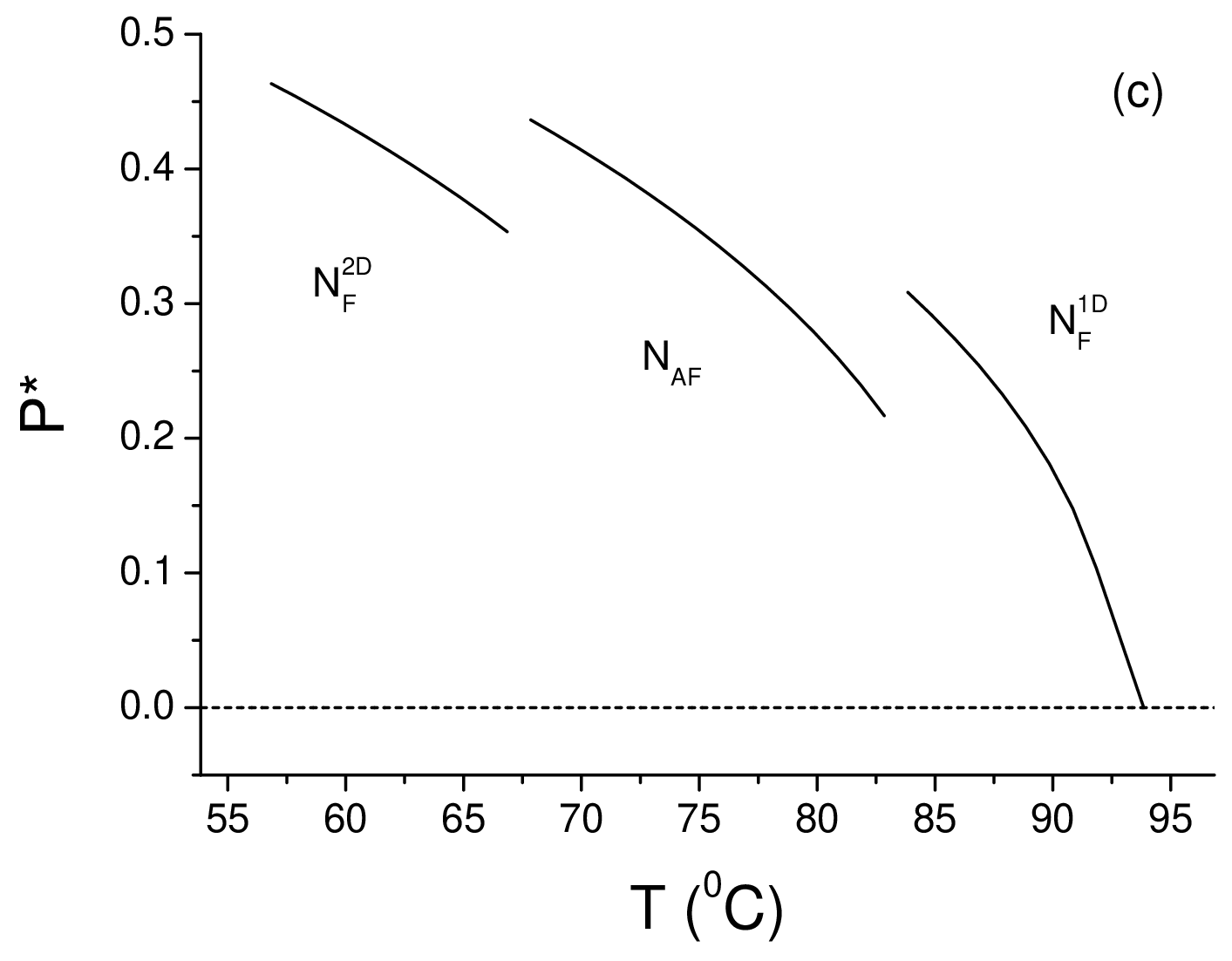}
\caption{\label{fig:epsart} Electric field -- temperature phase
diagram (a); Temperature dependencies of the domain radius (b) and
characteristic polar order parameter (c) at $E=0$. Green arrow in
(a) tentatively corresponds to the phase sequence with temperature
variation at fixed $E\ne 0$. Red arrows with numbers in (a) and
(b) correspond to the temperatures of specific phase transitions
observed experimentally [Sec.~III~B]. Dash blue line in (b)
corresponds to the half-thickness of the cell.}
\end{figure}
\begin{figure}[t]
\includegraphics[width=0.35\linewidth,clip]{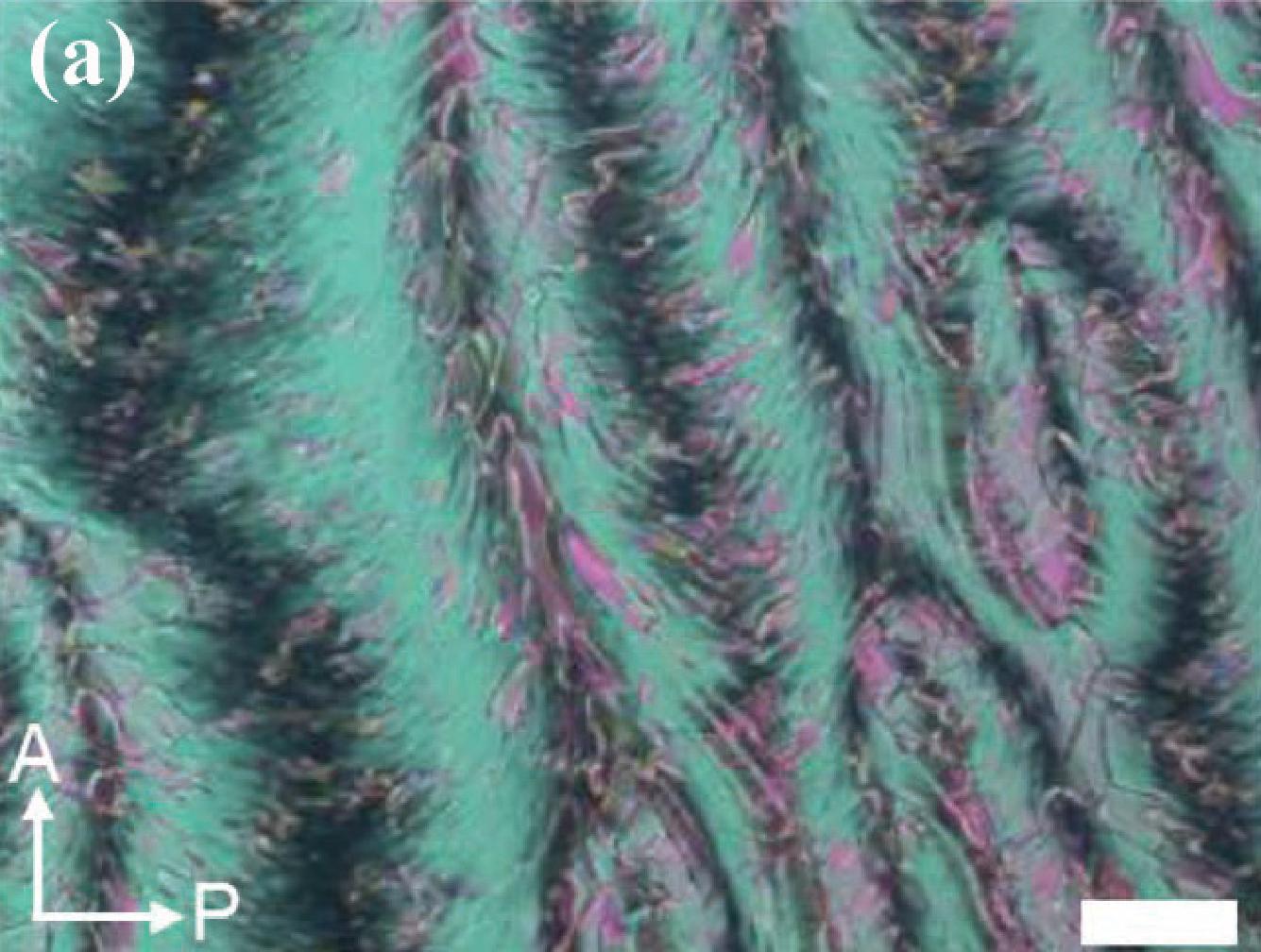}
\includegraphics[width=0.35\linewidth,clip]{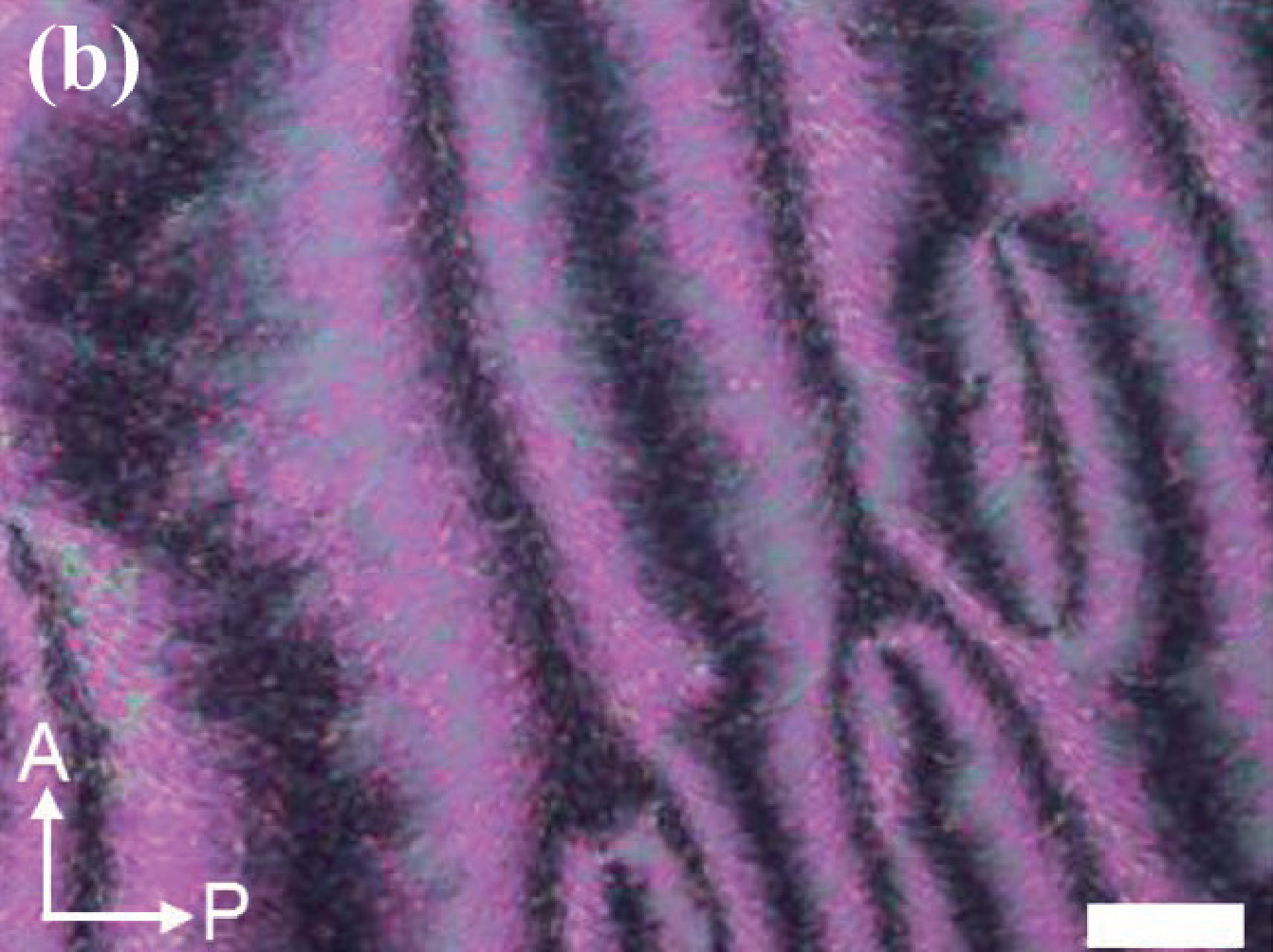}
\includegraphics[width=0.35\linewidth,clip]{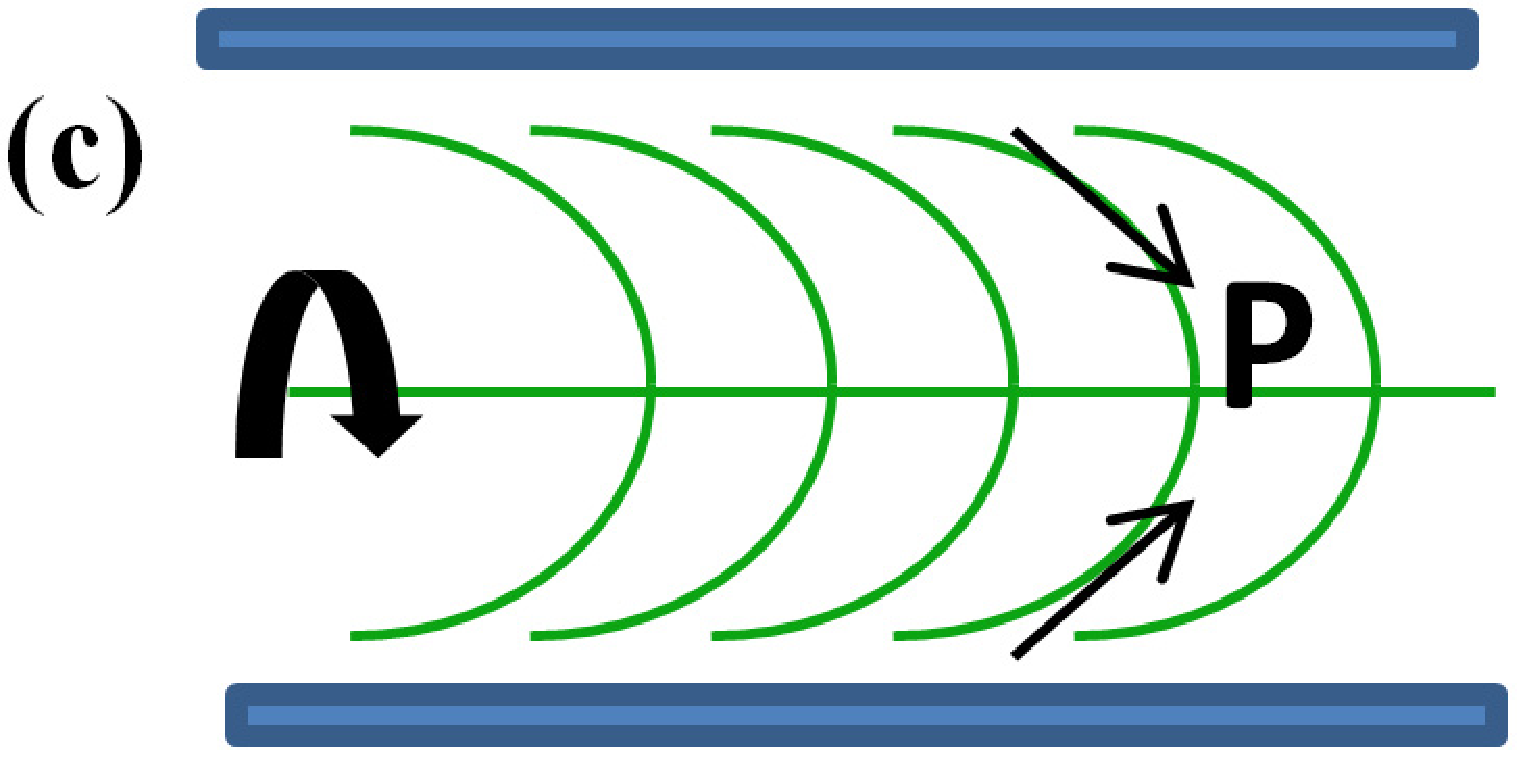}
\includegraphics[width=0.35\linewidth,clip]{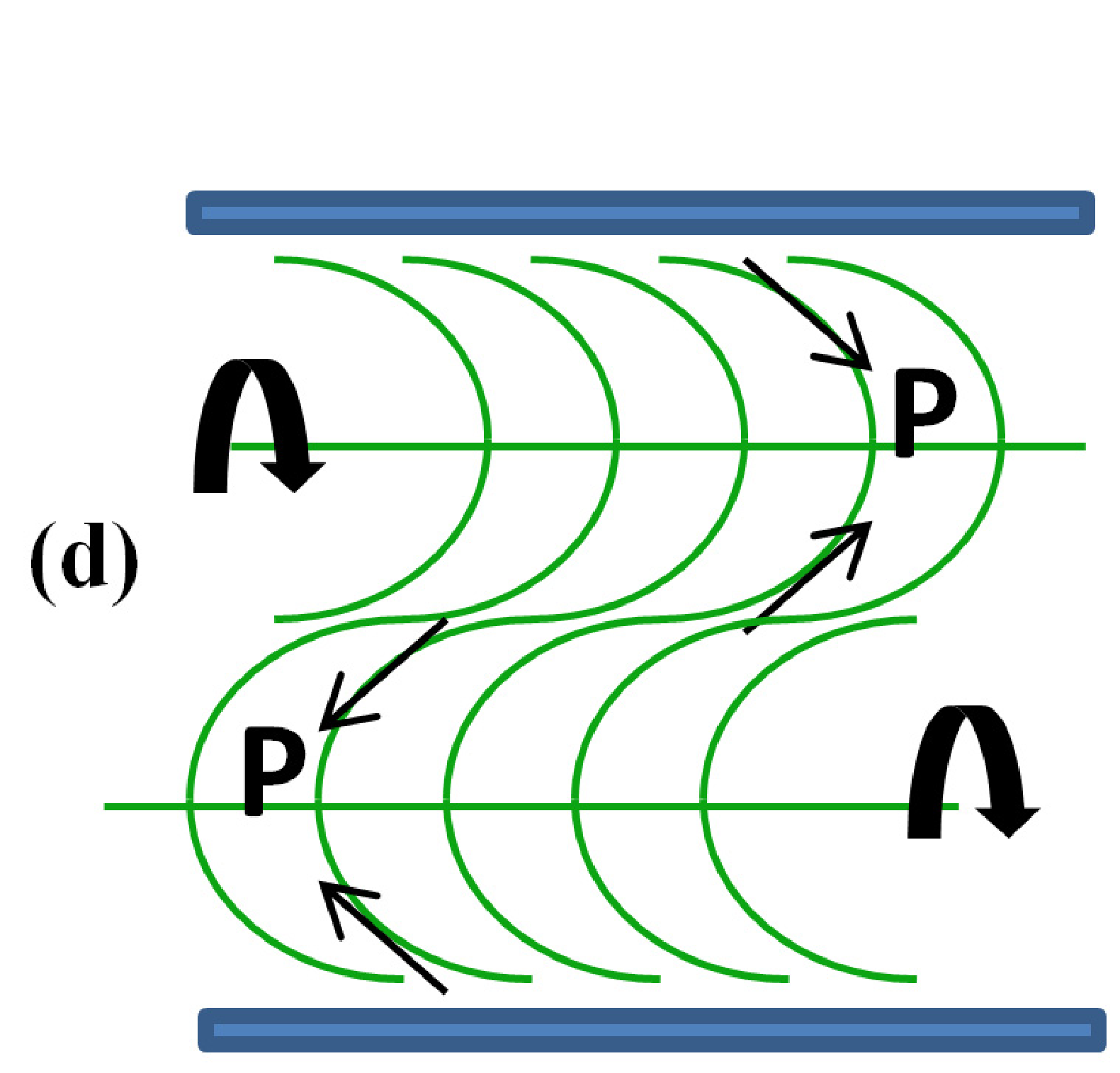}
\includegraphics[width=0.55\linewidth,clip]{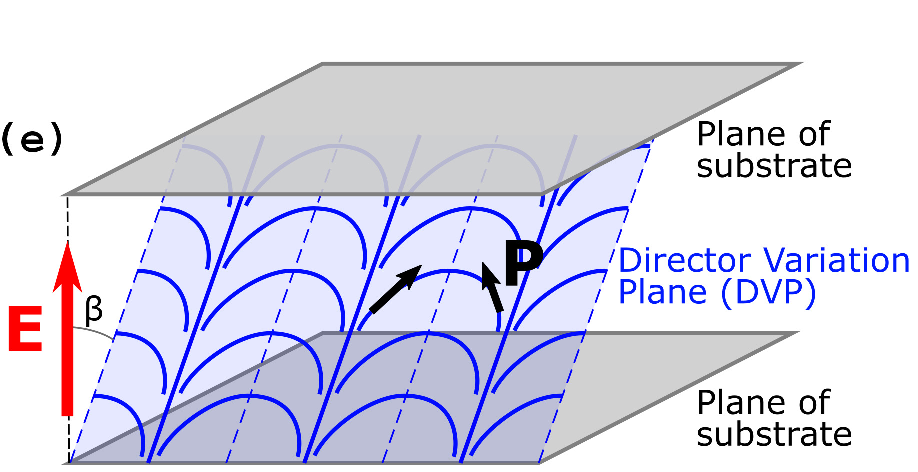}
\caption{\label{fig:epsart} POM images of quasi-cylindrical domains in $N_F^{2D}$ (a) and $N_{AF}$ (b); orientation of planar domains (schematic illustration) in $N_F^{1D}$ (c) and $N_{AF}^{1D}$ (d) at $E=0$; orientation of planar domains in $N_F^{1D}$ at $E\ne 0$. In (a) and (b) DIO material is used, cell thickness 10 $\mu$m, scale bar 100 $\mu$m. Images of (a) and (b) are reproduced with permission from Ref.
\cite{Nishikawa:2017}. Copyright WILEY-VCH Verlag GmbH \& Co.
KGaA, Weinheim, 2017.}
\end{figure}
Theoretical part requires, however, some geometrical
simplification, such as consideration of the axial and planar
symmetries. Within these symmetry restrictions, at various
conditions we find the three basic structures, which are presented
in Fig.~2. At the same time, from computer simulations (Sec.~IV~C)
we can see the more detailed information about the transformations
between structures presented in Fig.~2, and the transient
structures have obviously more complex geometry. The first basic
structure is the double-splay antiferroelectric nematic phase
$N_{AF}$ (designated also $N_X$ or $M2$ elsewhere) with
alternating signs of the splay and polarization in space. The
other two basic structures, stable at different conditions, are
the double- and single-splay ferroelectric nematics, $N_F^{2D}$
and $N_F^{1D}$, respectively. $N_F^{2D}$ and $N_{AF}$ are composed
of quasi-cylindrical periodical domains, while $N_F^{1D}$ is
composed of planar periodical domains. For each structure
presented in Fig.~2, the ${\bf x}$-axis can be introduced, to
which the director is parallel in the middle of each domain. In
$N_F^{2D}$ and $N_{AF}$, the director exhibits variation along
radius ${\bf r}$ of cylinder, while in $N_F^{1D}$ the director
exhibits variation along single space direction (for uniformity of
equations, also designated as ${\bf r}$). In all cases, ${\bf r}$
is perpendicular to ${\bf x}$. In ferroelectric phases, $N_F^{2D}$
and $N_F^{1D}$, the projection of polarization on the ${\bf
x}$-axis does not alternate in sign, while in antiferroelectric
$N_{AF}$, polarization alternates periodically in sign along each
Cartesian coordinate.

The electric field -- temperature phase diagram is presented in
Fig.~3~(a), while the temperature dependencies of the domain
radius and characteristic polar order parameter at $E=0$ are
presented in Figs.~3~(b) and (c), respectively. From theory
(Sec.~IV~B) it follows that, within each polar phase, the domain
radius $r_m$ increases and polarization $P^*$ decreases with the
increasing temperature, while multiple $r_m\,P^*$ remains
constant. At zero electric field, $N_F^{2D}$ minimizes the free
energy at lower temperature, $N_F^{1D}$ phase minimizes the free
energy at higher temperature, and $N_{AF}$ minimizes the free
energy in between. The domains in $N_F^{2D}$ and $N_{AF}$ are
visible in microscope [see Figs.~4~(a) and (b), respectively],
their typical size is several micrometers. In the absence of
electric field, the domains in $N_F^{1D}$ at planar anchoring
[Fig.~4~(c)] are not visible, because the energy-optimal
configuration of the domains makes no optical difference between
any points on the glass substrate. In the absence of electric
field, the antiferroelectric single-splay phase [Fig.~4~(d)]
possesses the same free energy as the ferroelectric one. The plane
of each arc in Figs.~4~(c) and (d) can be vertical or tilted. At
moderate values of electric field, all the splay nematic phases
transform into $N_F^{1D}$ [the orientations of arcs in $N_F^{1D}$
in the presence of moderate electric field are shown in
Fig.~4~(e), they can be vertical ($\beta=0$) or tilted ($\beta\ne 0$) to fit the cell gap]
and at higher values of electric field -- into paraelectric $N$
having uniform director orientation.
\begin{figure*}[tbh!]
\includegraphics[width=0.95\linewidth,clip]{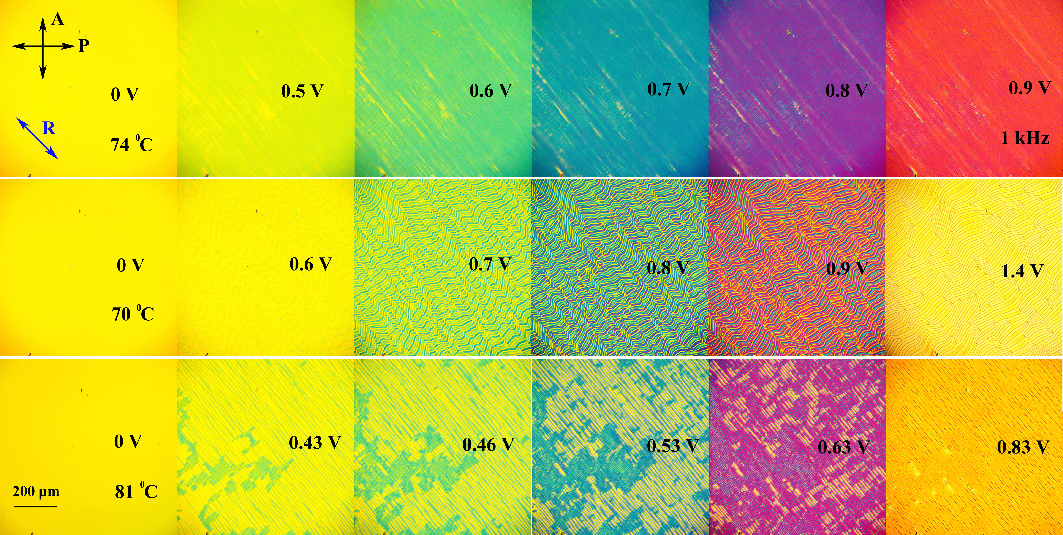}
\caption{\label{fig:epsart} POM
images of the DIO planar cell (cell thickness is 5 $\mu m$) at
several voltages (sinusoidal signal, $1\,kHz$) and temperature $T
= 74^{\circ}C$ (Row 1), $70^{\circ}C$ (Row 2) and $81^{\circ}C$
(Row 3). P and A are directions of polarizer and analyzer, R is
the rubbing direction.}
\end{figure*}

\section{Transformations of polar nematic phases induced by variation of temperature and electric field}
\subsection{Transformation of $N_{AF}$ and $N_F^{2D}$ into $N_F^{1D}$ in electric field}

Let us consider the variation of $N_{AF}$ in the electric field,
as it seen from POM and computer simulations (Sec.~IV~C). The
$5\mu m$-thick planar cell was filled with DIO liquid crystal in
isotropic state. After that, the sample was examined by polarizing
optical microscope (Nikon V100N Pol, Japan) equipped with a
heating stage (TMS-93 Stage Temp Controller and THMS 600
microscope stage, UK). The voltage from a waveform generator
(Agilent 33220A, USA) was applied to the ITO-coated cell
substrates.

On cooling in the absence of electric field, $N_{AF}$ is observed
between $68.8^{\circ}C$ and $84.5^{\circ}C$ \cite{Nishikawa:2017}.
The $1\,kHz$ frequency electric field of various amplitude was
applied at several temperatures within this range. The images of
the structure variation at application of electric field are
presented in Fig.~5. From computer simulations (Sec.~IV~C) if
follows that the antiferroelectric splay remains in the plane of
the substrate and gradually disappears when the voltage increases,
while the ferroelectric splay arises in the direction
perpendicular to the glass and gradually increases. Starting from
particular voltage, the stripes corresponding to the director
periodical modulation in space rotate from longitudinal (parallel
to the rubbing direction) to transverse (perpendicular to the
rubbing direction).  In the $5\mu m$-thick planar cell, in
correspondence with Fig.~3~(b), the domains are bigger than the
cell thickness almost in the whole temperature range of $N_{AF}$
and are therefore suppressed at the ground state (at $E=0$) by the
surfaces. From birefringence measurements in
Ref.~\cite{Emelyanenko:2022} we also conclude that, at $0 V$, the
conventional paraelectric nematic phase is observed. Temperature
$74^{\circ}C$ (Row~1 in Fig.~5) corresponds to the middle of the
temperature range of $N_{AF}$ in the infinite bulk. At $0.6 V$ the
structure with longitudinal stripes, which is similar to the one
obtained in computer simulations, arises. When the voltage
increases, one can observe the gradual disappearance of the
longitudinal stripes and appearance of the transverse ones.  At
$70^{\circ}C$ (Row 2 in Fig.~5) the AF to F transition threshold
[Fig.~3~(a)] corresponds to the lower voltage, therefore $0.6 V$
is sufficiently large voltage to cross over directly to the
ferroelectric state, and only the transverse stripes are observed.
One notes, that at each transverse stripe, the middle of each arc
presented in Fig.~4~(e) fully satisfies the planar alignment at
vertical orientation of director variation plane ($\beta=0$). When the voltage
increases, the director modulation first increases, but at higher
voltage starts decreasing again. At $81^{\circ}C$ (Row 3 in
Fig.~5) the situation is generally the same. First, the
longitudinal stripes appear, then the transverse ones.
Surprisingly, at any temperature, at higher voltage the structure
is targeted to become planar paraelectric again. This mainly
happens due to disbalance between the induced and flexoelectric
polarizations, which is discussed in details in Sec.~IV~B.

\begin{figure}[h!]
\includegraphics[width=0.32\linewidth,clip]{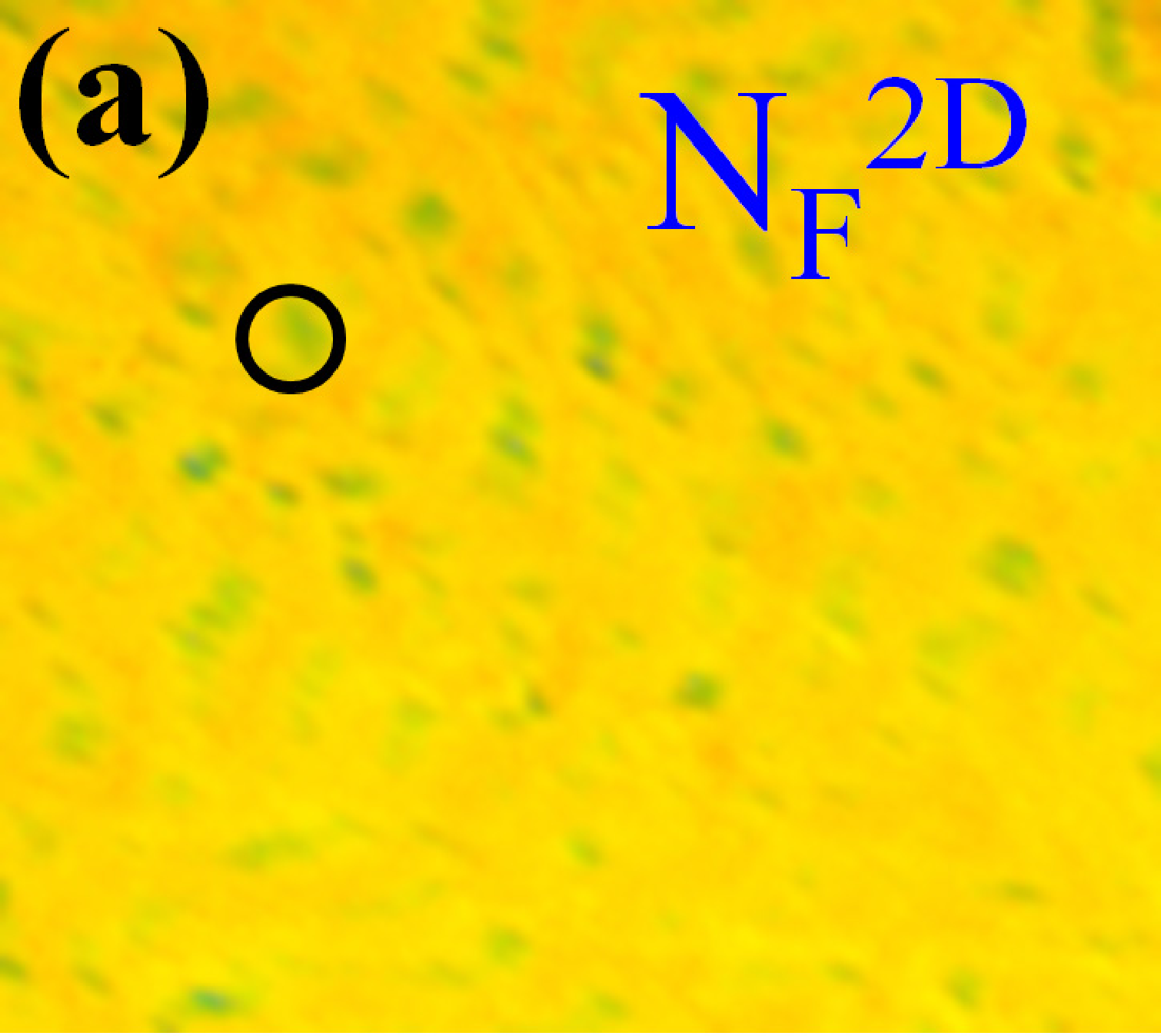}
\includegraphics[width=0.28\linewidth,clip]{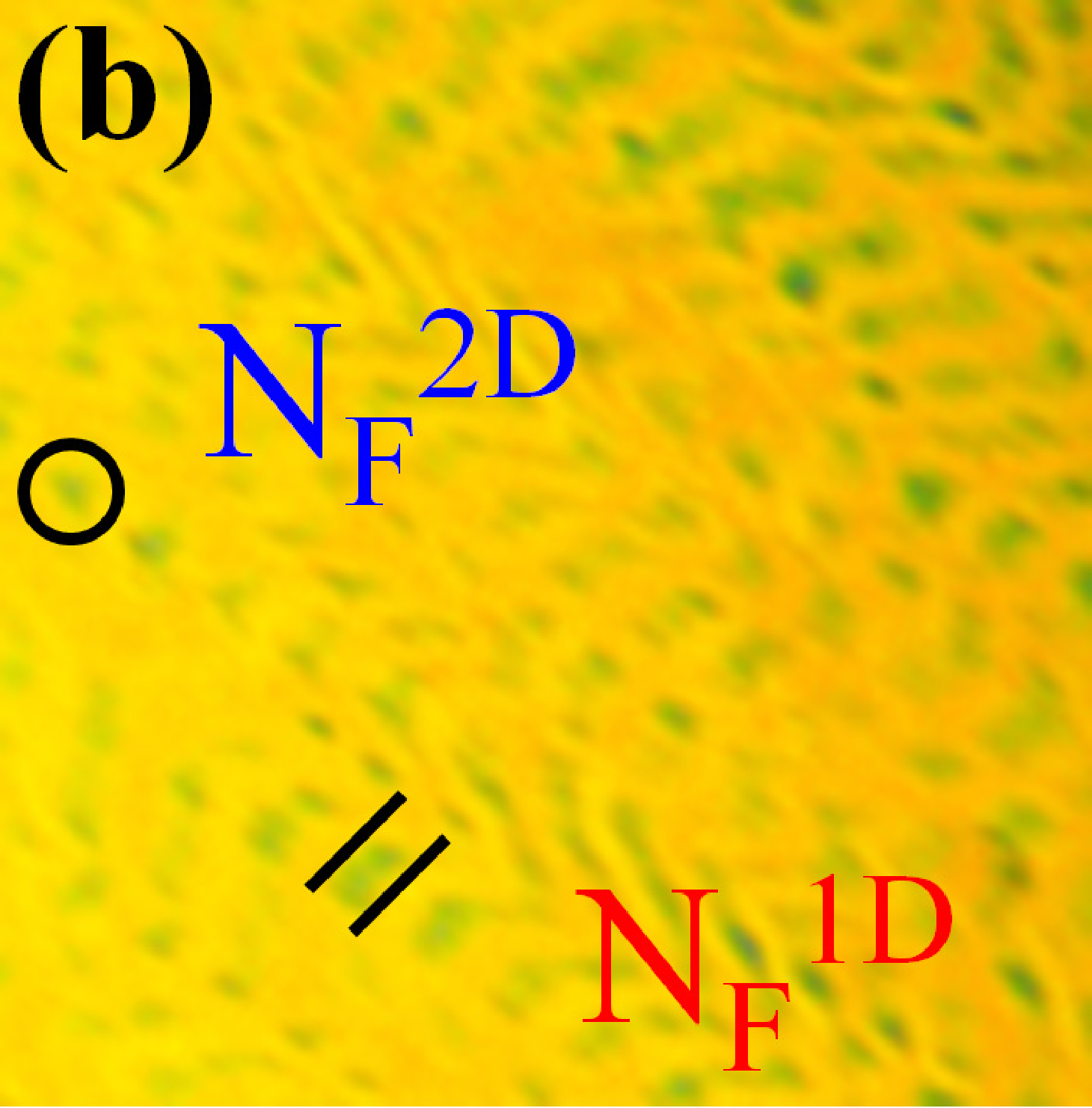}
\caption{\label{fig:epsart} POM images of the two parts of the
cell with DIO at $67^{\circ}C$ under $0.4\,V$, $1\,kHz$ electric
field: (a) mostly $N_F^{2D}$; (b) transformation from $N_F^{2D}$
to $N_F^{1D}$.}
\end{figure}

From experimental observation it also follows that transformation
from $N_F^{2D}$ to $N_F^{1D}$ induced by electric field also
happens continuously, while all the phase borders presented in
Fig.~3~(a) follow from consideration of simplified geometries
(either planar or cylindrical) and rather indicate tentative
places on the diagram where the continuous transformations between
phases should happen. In Fig.~6, a POM images of a planar cell of
DIO material at $67^{\circ}C$ (just below the temperature of
transition from $N_F^{2D}$ to $N_{AF}$) are presented at
application of $0.4\,V$, $1\,kHz$ electric field. The major part
of the cell [Fig.~6~(a)] represents a conjugation of $N_F^{2D}$
cylindrical domains with each other, similar to those presented in
Fig.~3~(b). Another part of the cell [Fig.~6~(b)] demonstrates a
continuous transformation from $N_F^{2D}$ to $N_F^{1D}$. The
quasi-cylindrical domains continuously transform into the
elongated ones and then to the linear stripes. Here the darker
dots and lines correspond to $\theta\rightarrow 0$ and $\pi$  (the
places where the director is parallel to electric field -- at
cylinder axes or in the middle planes of planar domains, see
definition of angle $\theta$ in Fig.~2). The brighter surrounding
corresponds to $\theta\rightarrow \pi/2$ (the places where the
director is parallel to the substrate -- at the domain periphery).
At higher voltage, the whole system transforms into $N_F^{1D}$ and
then into paraelectric $N$.

\subsection{Variation of the structure of DIO on cooling at applied voltage}

Let us consider the temperature-induced phase transitions in polar
nematic in the presence of electric field. The $5\mu m$-thick
planar cell filled with DIO material was cooling from
$99^{\circ}C$ down to $27^{\circ}C$ at applied $1 kHz$ frequency
electric field with constant amplitude $0.9 V$. The images in
crossed polarizers were registered each half-degree. The phase sequence generally appears to be
completely different from that observed without electric field.
Particular key images are presented in Figs.~7 and 8. The images
practically do not change between $99^{\circ}C$ and $87^{\circ}C$
(see Fig.~7). Presumably we observe the uniform paraelectric
nematic with planar orientation of director in this temperature
range. One notes that our theoretical phase diagram presented in
Fig.~3~(a) predicts the existence of the $N_F^{1D}$ polar phase
below $93^{\circ}C$. However, in correspondence with Fig.~3~(b),
an equilibrium domain length (corresponding to the infinite bulk
of LC) in the temperature range between $93^{\circ}C$ and
$87^{\circ}C$ is predicted to be huge, and, in realistic confined
geometry, the splay domains are most likely suppressed by the
substrates. However, below $87^{\circ}C$ the images start
gradually becoming darker, and some longitudinal stripes arise.

The red arrows with numbers presented in Figs.~3~(a) and (b)
reflect the temperatures of particular phase transitions observed
in DIO at applied fixed voltage, and Arrow 1 tentatively
corresponds to the realistic temperature of transition from
paraelectric $N$ to $N_F^{1D}$ in DIO confined between parallel
glasses ($87^{\circ}C$).  At $87^{\circ}C$ the equilibrium domain
size in $N_F^{1D}$ is still greater than the cell thickness.
However, we expect that the highly tilted (almost parallel to the
substrate, $\beta\rightarrow\pi/2$) single-splay domains can already exist. The electric
field would like to make the director variation plane vertical
[perpendicular to the substrates, $\beta=0$ in Fig.~4~(e)], but in this
case the splay domains would not fit the gap between glasses. In
this situation, both arms of each arc in Fig.~4~(e) choose the
longitudinal (parallel to the rubbing) orientation, and this could
be the origin of the longitudinal stripes observed in the
temperature range between $87^{\circ}C$ and $82^{\circ}C$. When
the temperature decreases down to $82^{\circ}C$, the domain size
decreases continuously [see Fig.~3~(b)], therefore the arcs
presented in Fig.~4~(e) gain larger and larger vertical projection
(which is favorable for the coupling of flexoelectric polarization
with electric field), and the images in crossed polarizers are
becoming darker and darker.

Surprisingly, the structure variation does not demonstrate any
irregular variation near the transition into $N_{AF}$ registered
at $84.5^{\circ}C$ in Ref.~\cite{Nishikawa:2017} by DSC
measurements in the absence of electric field, which indirectly
indicates that the structure of LC does not have any tendency to
return to the ground state between pulses of high-frequency
electric field. An easy explanation for this effect is that the
switch-off relaxation time should be much longer than the inverse
frequency of electric field in this temperature range. This could
be related to the fact that flexoelectric polarization inversion
requires the director splay inversion in the whole space. However,
the director is trapped by its own periodical structure. The
alternative to the director continuous motion is the total
director disruption in the whole space, whose energy cost is much
higher. Since the director is determined on the scale, which is
much larger than molecular size, the director motion is analogous
to that of the Brownian particles, whose velocity is much slower
than molecular velocity. Our expectation following from
Einstein-Smoluchowski equation is that director should not move
faster than several micrometers per second, which means that
application of $1 kHz$ frequency electric field should definitely
trap the director distribution within the 5 $\mu m$-thick cell,
since the director can only move a few nanometers per a pulse. If
the frequency of electric field is between the inverse $\tau_{\rm
on}$ and $\tau_{\rm off}$, the director should stay in the
position of particular (let's say, positive) pulse, and should not
return to the ground state between pulses. Detailed analysis of
all the images of DIO material obtained at application of the
high-frequency electric field demonstrates good correlation with
our theoretical predictions obtained in a supposition of constant
electric field used in Sec.~IV.

\begin{figure}[h!]
\includegraphics[width=0.7\linewidth,clip]{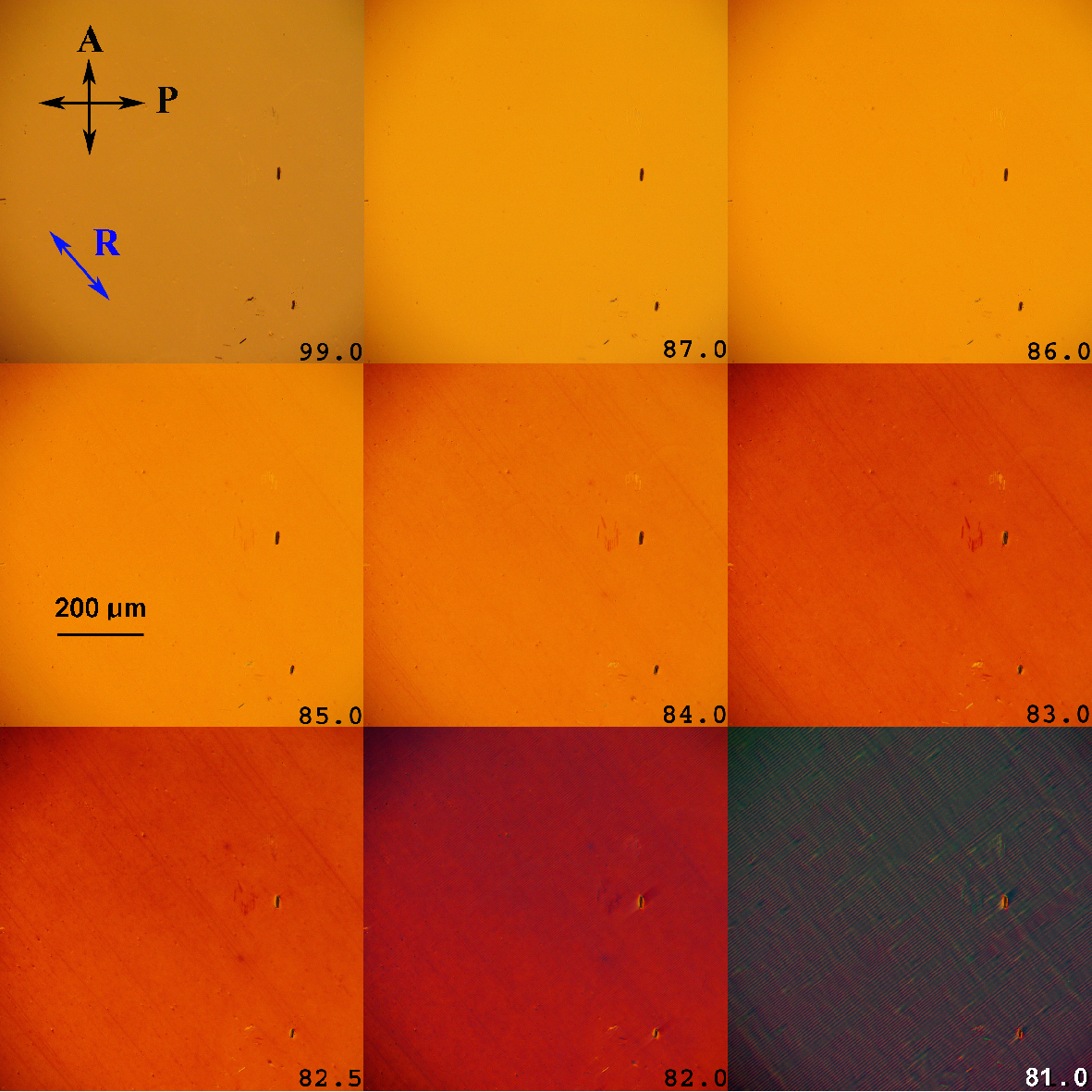}
\caption{\label{fig:epsart} POM images of the DIO planar cell
(cell thickness is 5 $\mu m$) at applied fixed electric field
($0.9 V$, sinusoidal signal, $1\,kHz$) during the cooling cycle
(temperatures are indicated in the bottom right corners). P and A
are directions of polarizer and analyzer, R is the rubbing
direction.}
\end{figure}

\begin{figure}[h!]
\includegraphics[width=0.7\linewidth,clip]{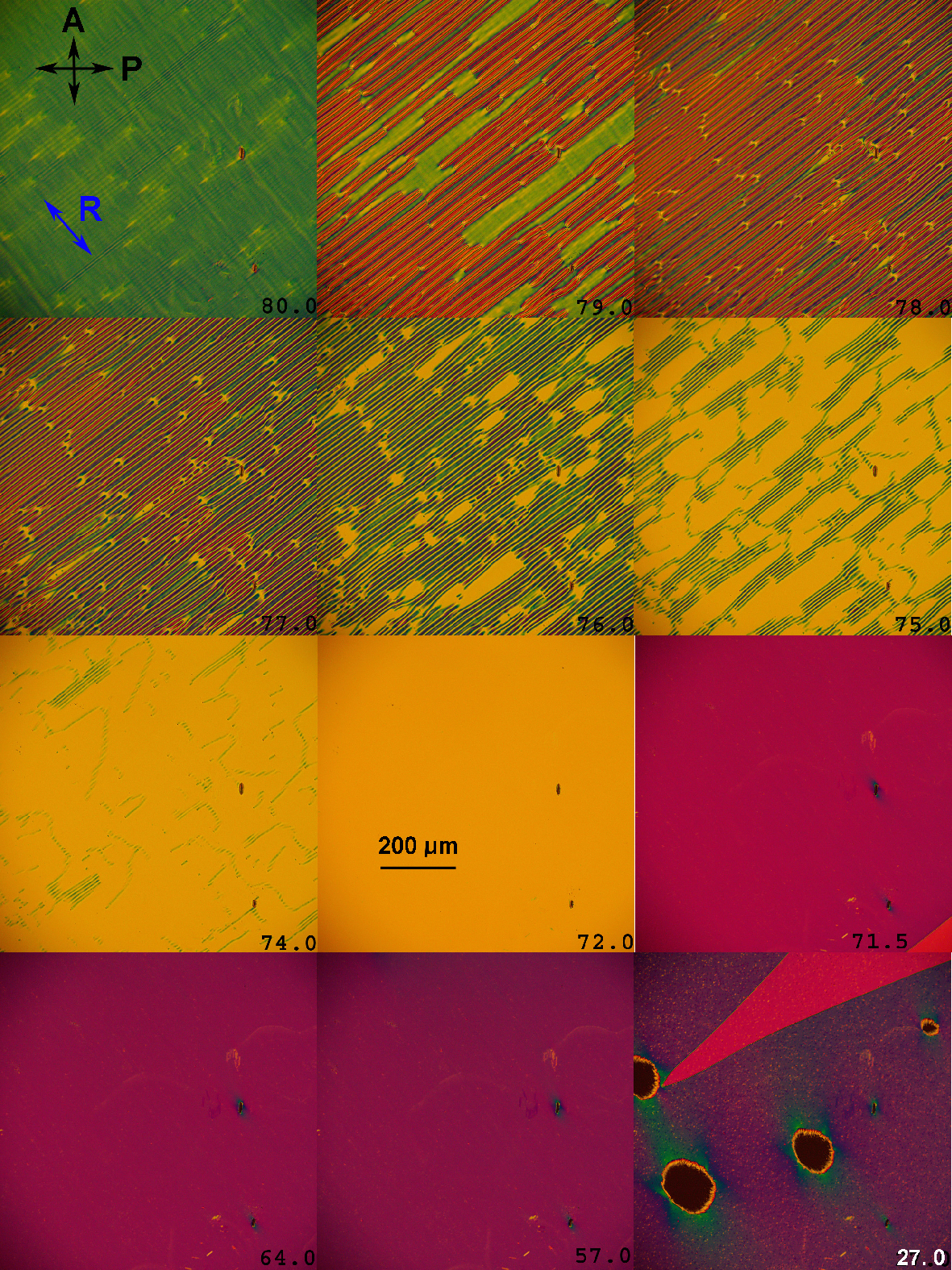}
\caption{\label{fig:epsart} POM images of the DIO planar cell
(cell thickness is 5 $\mu m$) at applied fixed electric field
($0.9 V$, sinusoidal signal, $1\,kHz$) during the cooling cycle
(continuation of Fig.~7).}
\end{figure}

The next phase transition [marked with Arrow 2 in Fig.~3~(b)]
takes place at some temperature between  $82^{\circ}C$ and
$81^{\circ}C$ (see Fig.~7), at which the transverse (perpendicular
to the rubbing direction) stripes appear, and the whole image
becomes darker step-wise. This transition is most likely the
first-order and is related to the reorientation of director
variation plane [reorientation of each arc presented in Fig.~4~(e)
around vertical axis]. The reason is that the vertical projection
of each arc is already sufficiently great at $82^{\circ}C$, and
the middle of each arc is targeted to be oriented along the
rubbing direction, while the arms of each arc, on the contrary,
are not biased anymore.

From observation of images presented in Fig.~8 it follows, that
between $80^{\circ}C$ and $79^{\circ}C$ [Arrow 3 in Fig.~3~(a)]
the transition from $N_F^{1D}$ to the phase corresponding to
$N_{AF}$ deformed in the electric field (see discussion in
Sec.~III~A) takes place. Tentatively, the structure variation in
this temperature range follows the green arrow in Fig~3~(a), which
is inclined, because the splay elastic constant (participating in
the denominator of the value plotted on the vertical axis) should
tentatively increase with the decreasing temperature, at least
within the range of a single phase. One can see the islands of
$N_{AF}$ inside of $N_F^{1D}$ at $80^{\circ}C$, and therefore the
$N_F^{1D}$ to deformed $N_{AF}$ phase transition is also of the
first order.

At about $74^{\circ}C$ the structure comes out of the $N_{AF}$
range [Arrow 4 in Fig.~3~(a)]. Formally, the structure should
return to $N_F^{1D}$. However, it was demonstrated in Sec.~III~A
that, at this temperature and voltage, the director splay modulation is
not very deep, and the structure rather resembles the paraelectric
nematic with planar director orientation. The origin of this
effect will be discussed Sec.~IV~B.

At $71.5^{\circ}C$ the new phase transition happens [Arrow 5 in
Fig.~3~(b)]. From our observations in Ref.~\cite{Emelyanenko:2022}
and also from Fig.~3~(b) it follows that the size of domains in
$N_{AF}$ becomes comparable to the cell thickness at around
$71.5^{\circ}C$, and thus, the stripes corresponding to the splay
domains in $N_{AF}$ would arise at $E=0$. Above $71.5^{\circ}C$
the structures of DIO at applied voltage and in the ground state
are the same -- the planar paraelectric nematic, while below
$71.5^{\circ}C$ they become different again. The observed
structure returns to the one resembling $N_F^{1D}$ observed
between $82^{\circ}C$ and $87^{\circ}C$ with partial inclusions of
the $N_F^{2D}$ domains, whose axes (visible as reflecting dots)
are oriented parallel to the electric field and perpendicular to
the substrates. When the temperature farther decreases (see
Fig.~8), the images become darker, and the number of reflecting
dots increases. The structure variation completely ignores the
$N_{AF}$ to $N_F^{2D}$ transition registered at $68.8^{\circ}C$ by
DSC measurements in Ref.~\cite{Nishikawa:2017} in the absence of
electric field, and thus, the structure does not return to the
ground state again, similarly to that in the temperature range between $82^{\circ}C$ and
$87^{\circ}C$.

At $57^{\circ}C$ the material structure is already close to the
nominal transition into $N_F^{2D}$. At $27^{\circ}C$, the
structure corresponding to the quasi-ideal $N_F^{2D}$ with islands
of crystal and also with some domains similar to those reported in
Ref.~\cite{Sebastian:2021}, consideration of which is beyond the
scope of the present paper, arises. At different temperatures we
observe similar domains at application of much higher voltage, at
which we already expect an induction of paraelectric nematic phase
by electric field (see discussion in Secs.~III~A and IV~B). One
notes that, at planar boundary conditions, in the presence of
electric field we obtained an image of $N_F^{2D}$ similar to that
obtained at homeotropic boundary conditions without electric field
in Ref.~\cite{Nishikawa:2017}.

\section{Theoretical approaches}
\subsection{Molecular-statistical theory: temperature and electric field dependent distributions of $S$ and $P$ order parameters in space}
Let us consider a system of elongated polar molecules (with
longitudinal electric dipoles {\mbox {\boldmath $\mu$}})
interacting with each other and with external electric field
${\bf E}$ (Fig.~9). Formally, constant electric field participates
in all equations below. Having in mind our discussion in Sec.
III~B about slow relaxation of director splay, we expect that the
structures arising at high-frequency electric field do not differ
very much from those obtained at constant electric field. In the
general case, director field ${\bf n}({\bf r})$ is
inhomogeneous, and the free-energy density $\partial F/\partial V$ can be written in the following form Ref.~\cite{Emelyanenko:2021}:
\begin{eqnarray}
4\pi V_0\frac{\partial F({\bf r}_1)}{\partial V}= k_B T\int d^2
{\bf a}_1 f[({\bf a}_1\cdot{\bf n}_1),{\bf r}_1] \ln f[({\bf
a}_1\cdot{\bf n}_1),{\bf r}_1] \nonumber\\ +\frac{\sigma_0}{8\pi
V_0}\int d^2 {\bf a}_1 \int d^2 {\bf a}_2 \int d^3 {\bf r}_{12}
f[({\bf a}_1\cdot{\bf n}_1),{\bf r}_1]
 f[({\bf a}_2\cdot{\bf n}_2),{\bf r}_2]
U_{12}^{ef}({\bf a}_1,{\bf a}_2,{\bf r}_{12}) \nonumber\\ -4\pi\mu(\sigma_0+1)\int
d^2 {\bf a}_1 f[({\bf a}_1\cdot{\bf n}_1),{\bf r}_1]({\bf
a}_1\cdot{\bf E})\quad,\quad
\end{eqnarray}
where $V_0$ is the bulk occupied by a
molecule located at point ${\bf r}_1$ and all its nearest
neighbors, $\sigma_0$ is the average number of neighbors for each
molecule, $f[({\bf a}\cdot{\bf n}),{\bf r}]$ is the orientational
distribution function for molecules having principal axes ${\bf
a}$ with respect to director ${\bf n}$ at point ${\bf r}$,  ${\bf
r}_i$ ($i=1,2$) are the coordinates of points $1$  and $2$, where
molecules $1$ and $2$ are located, ${\bf r}_{12}$ is the vector
connecting points $1$ and $2$, $k_B$ is the Boltzmann constant,
$T$ is the temperature, $U_{12}^{ef}({\bf a}_1,{\bf a}_2,{\bf
r}_{12})$ is the effective pair interaction potential for two
molecules with long axes ${\bf a}_1$ and ${\bf a}_2$ located at
points $1$ and $2$, respectively, while ${\bf n}_1$ is the
director at point $1$ and ${\bf n}_2$ is the director at point
$2$. The first term in Eq.~(5) is the entropy, the second term is
the internal energy, and the third term is the energy of
interaction of molecular longitudinal dipoles with electric field.
At any point ${\bf r}$, the orientational
distribution function $f[({\bf a}\cdot{\bf n}),{\bf r}]$ in
Eq.~(5) satisfies the normalizing constraint:
\begin{equation}
\int d^2 {\bf a} f[({\bf a}\cdot{\bf n}({\bf r })),{\bf r}]=1
\quad. \quad
\end{equation}
\begin{figure}[h!]
\includegraphics[width=0.4\linewidth,clip]{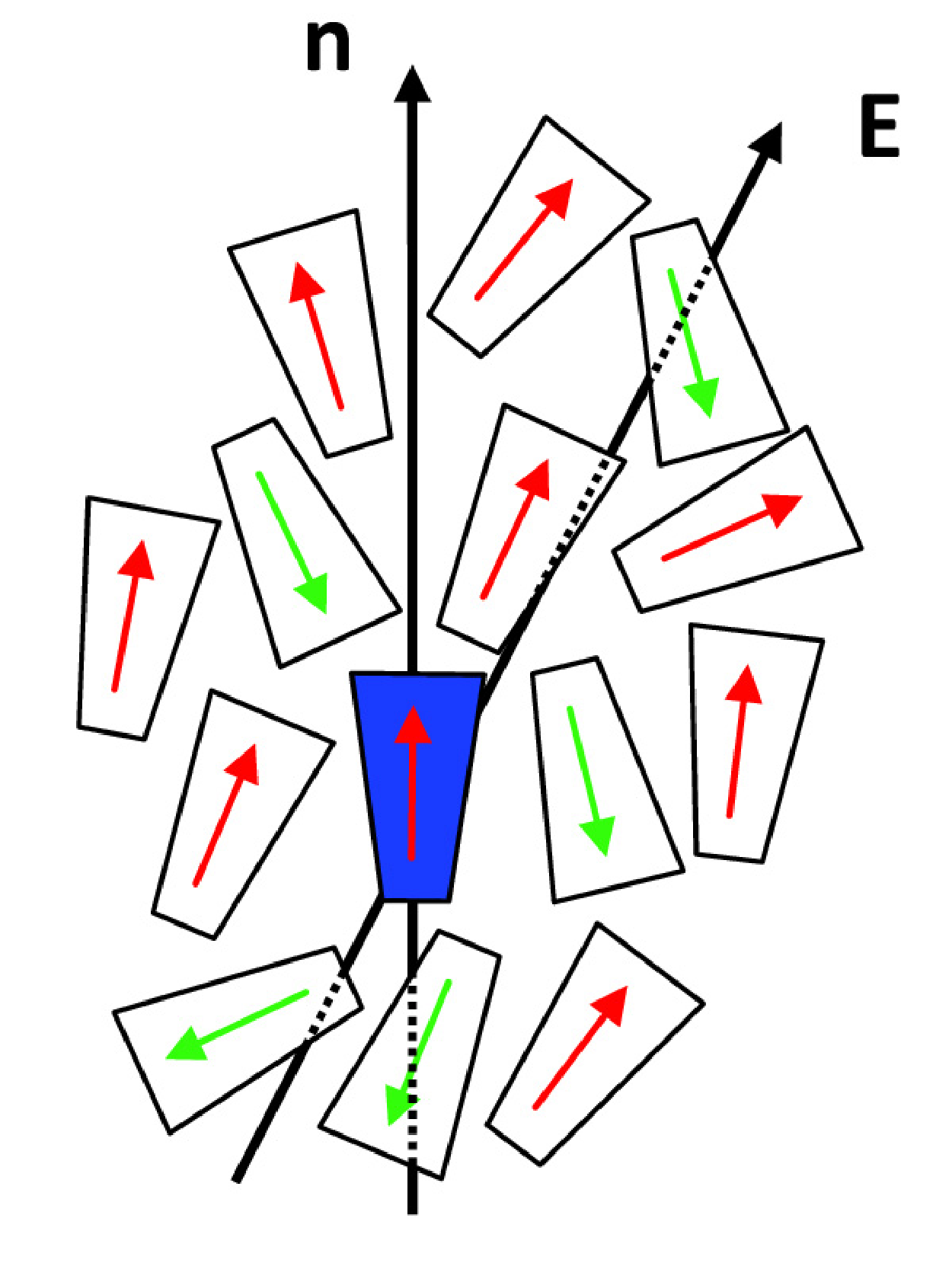}
\caption{\label{fig:epsart} A trial (blue) polar molecule in a
combination of the mean molecular field and electric field ${\bf E
}$. Here ${\bf n}$ is the local director at a point, where trial
molecule is located.}
\end{figure}
Minimizing the free energy (5) with respect to orientational
distribution function $f[({\bf a}\cdot{\bf n}),{\bf r}]$ under
constraint (6), one obtains:
\begin{eqnarray}
f[({\bf a}\cdot{\bf n}),{\bf r}]=\frac{1}{I_0({\bf r})}
\exp\biggl\{-\frac{U_{MF+E}[({\bf a}\cdot{\bf n}),{\bf r}]}{k_B
T}\biggr\} \quad,\quad\nonumber\\
\end{eqnarray}
where ${\bf r}\equiv{\bf r}_1$, ${\bf n}\equiv{\bf n}_1$,
$I_0({\bf r})$ is the normalizing constant, and $U_{MF+E}[({\bf
a}\cdot{\bf n}),{\bf r}]$ is the potential of a molecule located
at point ${\bf r}$ with orientation ${\bf a}\equiv{\bf a}_1$ of
prime axis in a combination of the mean molecular field and
electric field:
\begin{eqnarray}
U_{MF+E}[({\bf a}\cdot{\bf n}),{\bf r}] \equiv
\frac{\sigma_0}{4\pi V_0}\int d^3 {\bf r}_{12} \int d^2 {\bf a}_2
f[({\bf a}_2\cdot{\bf n}_2),{\bf r}_2]
\biggl[U_{12}^{ef}({\bf a}_1,{\bf a}_2,{\bf r}_{12}) -\mu ({\bf
a}\cdot{\bf n})({\bf n}\cdot{\bf E})\biggr]
 \quad.\quad
\end{eqnarray}
Approximating the pair potential by spherical invariants
[Eq.~(1)], substituting Eq.~(1) into Eq.~(8), introducing
coefficients
\begin{equation}
J_{\ell L \lambda}^{(i)} \equiv \frac{\sigma_0}{4\pi
V_0}\int\limits_0^\infty d r_{12} r_{12}^{i+2} J_{\ell L
\lambda}(r_{12})
\end{equation}
and using only $T_{101}$, $T_{110}$, $T_{011}$ and $T_{202}$
spherical invariants resulting in average in the appearance of the terms in the mean field depending on the powers of operator ${\mbox{\boldmath $\nabla$}}$ not higher than one, one finally obtains the following expression
for the potential of a molecule with orientation ${\bf a}$
affected by a combination of the mean molecular field and electric
field:
\begin{eqnarray}
-U_{MF+E}(t,{\bf r})= J_{101}^{(0)}P({\bf r}) P_1(t) +
J_{202}^{(0)} S({\bf r}) P_2(t)
+\biggl\{\frac{1}{6}\biggl[J_{110}^{(1)}+J_{011}^{(1)}\biggr]({\mbox
{\boldmath $\nabla$}}\cdot{\bf n})+\mu({\bf n}\cdot{\bf
E})\biggr\}P_1(t) \quad,\quad
\end{eqnarray}
where $t\equiv({\bf a}\cdot{\bf n})$, $P_1(t)\equiv t$ and
$P_2(t)\equiv 3/2\,t^2-1/2$ are the first and the second Legendre
polynomials. Eq.~(10) corresponds to the first (simplest) approximation reflecting modulation of the $S$ and $P$ order parameters caused by modulation of splay and describing the major tendency: both parameters $S$ and $P$ should be higher (lower) at the places where the splay is higher (lower). The first two terms in Eq.~(10) are the polar and
non-polar anisotropies, while the two terms in figure brackets are
due to the flexoelectric effect and electric field. From Eqs.~(2),
(7) and (10) one readily obtains the following recurrent equations
for determination of the $P({\bf r})$ and $S({\bf r})$ order
parameters at each temperature $T$ and electric field $E$ at any
given ${\bf n}({\bf r})$ distribution:
\begin{eqnarray}
P({\bf r})=\frac{I_1({\bf r})}{I_0({\bf r})} \quad,\quad S({\bf
r})=\frac{I_2({\bf r})}{I_0({\bf r})} \quad,\quad
\end{eqnarray}
where integrals $I_m({\bf r})$ are defined as follows:
\begin{eqnarray}
I_m({\bf r})\equiv\int\limits_{-1}^{1} P_m(t)
\exp\biggl\{-\frac{U_{MF+E}(t,{\bf r})}{k_B T}\biggr\} d t
\quad,\quad
\end{eqnarray}
where $U_{MF+E}$ is determined by Eq.~(10). Substituting solution
(7)--(10) back into Eq.~(5), one obtains for the equilibrium
free-energy density $\partial F_{\rm eq}/\partial V$:
\begin{equation}
4\pi V_0\frac{\partial F_{\rm eq}({\bf r})}{\partial V}=-k_B T\ln I_0({\bf
r })+\frac{1}{2}J_{101}^{(0)}P^2({\bf r}) +
\frac{1}{2}J_{202}^{(0)} S^2({\bf r}),
\end{equation}
where normalizing integral $I_0({\bf r})$ should be calculated
using Eqs.~(12) and (10). Eq.~(13) should be used for comparison
of the free energies of the neighboring phases in the phase
diagram.

As it was mentioned in Sec. II,  Eq.~(4) [multiplied by $4\pi V_0$] is the explicitly depending on director ${\bf n}$ part of Eq.~(5). Indeed, if one prolongs the gradient expansion in Eq.~(10) up to the terms depending on the second power of operator ${\mbox{\boldmath $\nabla$}}$ (for this purpose, invariants $T_{220}$, $T_{022}$,  $T_{222}$,  $T_{422}$ and $T_{224}$ should also be considered in approximation Eq.~(1); this is done in Ref.\cite{Emelyanenko:2021}) and substitutes Eq.~(10) into the second and third terms of Eq.~(5) [definition Eq.~(8) should also be used], then one obtains Eq.~(4). In
particular, the flexoelectric and electric-field-dependent terms
(which explicitly depend on both $P$ and ${\bf n}$) coincide in
Eqs.~(4) and (5) at substitution of $4\pi V_0
K_{11}\lambda=[J_{110}^{(1)}+J_{011}^{(1)}]/6$ and $V_0
\varepsilon_a=\mu(\sigma_0+1)$. By the same substitution of Eq~(10) into Eq.~(5), one obtains the $P^2$ term introduced phenomenologically in Ref.~\cite{Sebastian:2020}. In the same manner, it is possible to obtain additional contributions to the elastic constants depending on the polar order parameter $P$ studied phenomenologically in Refs.~\cite{Mertelj:2018,Sebastian:2020} (see also Ref.~\cite{Emelyanenko:2022}).

\subsection{Perturbation elastic continuum theory reflecting space variation of $S$ and $P$ order parameters}

Let us now consider the director distribution in the polar phases
in the presence of electric field, having in mind the results
obtained in Sec.~IV~A. In the cases presented in Figs.~2~(a) and
(b), we expect that director is (mostly) along the radial planes
(the planes parallel to the cylinder axis ${\bf x}$ and radius
${\bf r}$). Thus, in all cases presented in Fig.~2, the director
mainly has two nonzero coordinates, similarly to that in
Ref.~\cite{Garbovskiy:2020}:
\begin{figure}[h!]
\includegraphics[width=0.45\linewidth,clip]{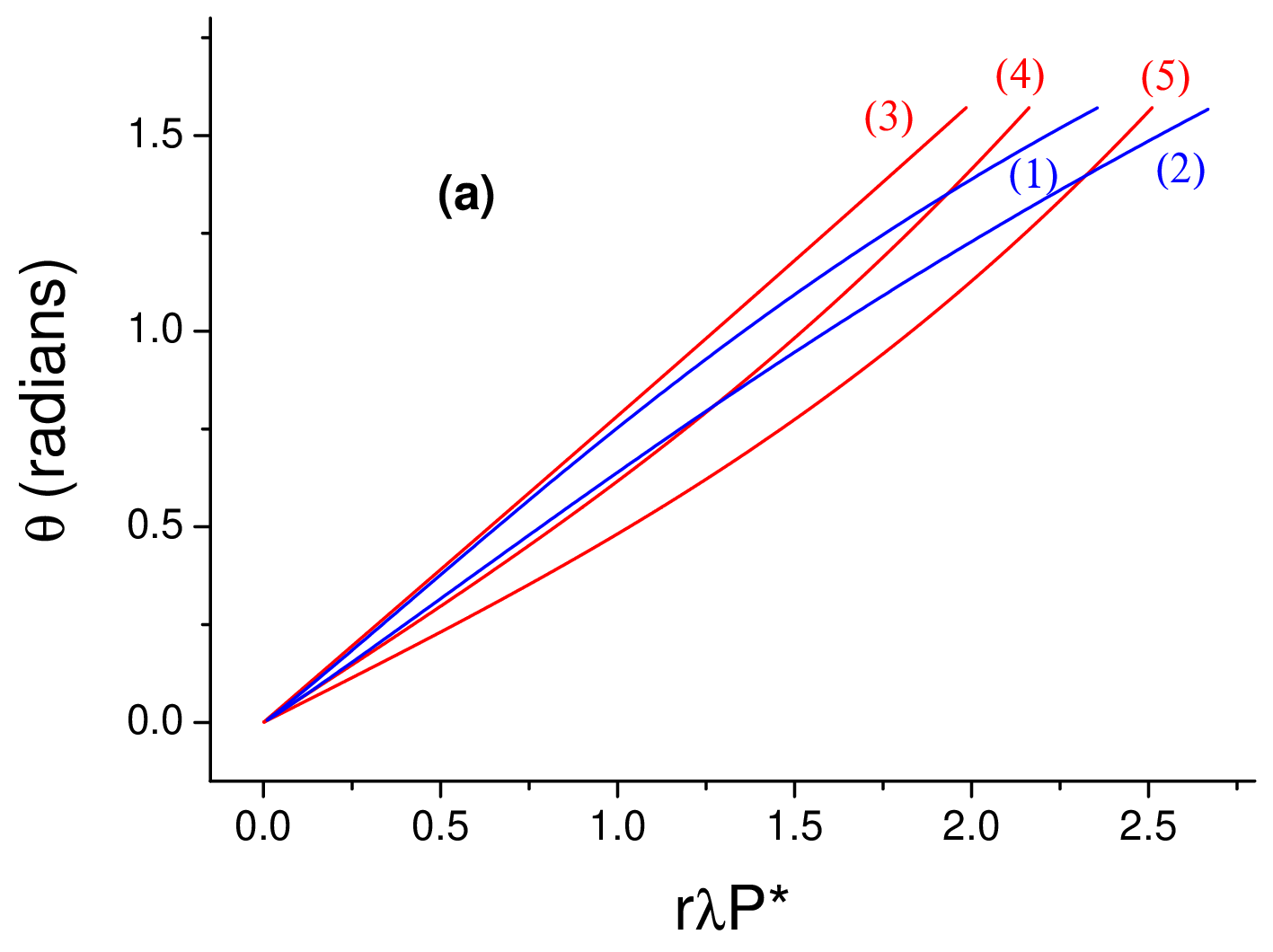}
\includegraphics[width=0.45\linewidth,clip]{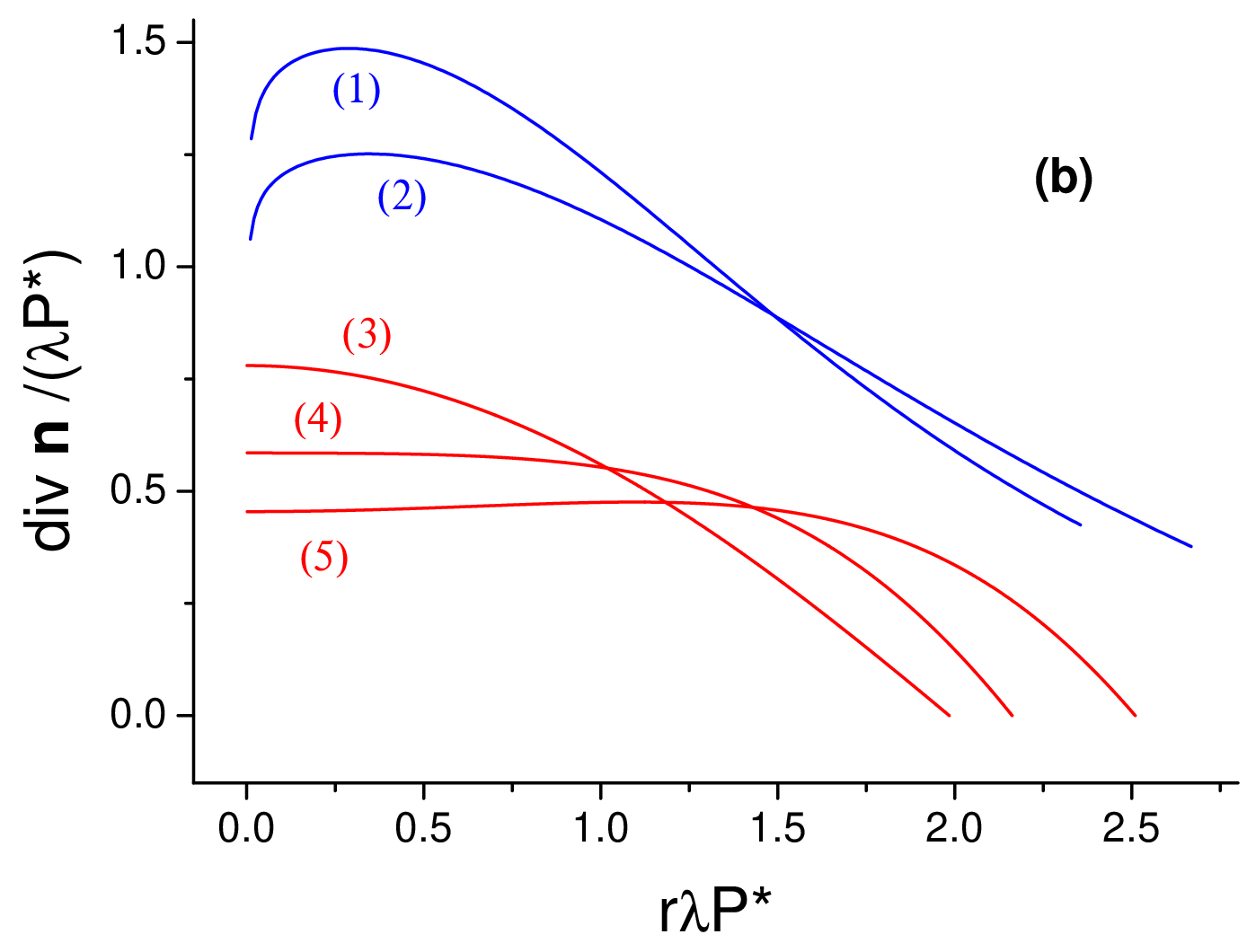}
\caption{\label{fig:epsart} Tilt of director (a) and splay
deformation (b) distributions within the domain in $N_F^{2D}$
[blue curves (1) and (2)] and $N_F^{1D}$ [red curves (3), (4) and
(5)]. Here $\tilde{\tau}=0.001$ (1), $0.22$ (2), $0.225$ (3),
$0.8$ (4) and $0.843$ (5).}
\end{figure}
\begin{equation}
n_x=\cos\theta(r) \quad,\quad  n_r=\sin\theta(r) \quad.\quad
\end{equation}
One notes, that all the structures presented in Fig.~2 can be
described in a unified way, and here we also introduce the
$\delta$ parameter to distinguish between the double- and
single-splay structures ($\delta$ is set to one in the case of
2D-splay and is set to zero in the case of 1D-splay). From
Eq.~(14) it follows:
\begin{eqnarray}
({\mbox {\boldmath $\nabla$}}\cdot{\bf
n})=\frac{\delta}{r}\sin\theta+\cos\theta\,\frac{d \theta}{d r}
\quad,\quad
[{\bf n}\times[{\mbox{\boldmath $\nabla$}}\times{\bf
n}]]^2=\sin^2\theta \biggl(\frac{d \theta}{d r}\biggr)^2
\quad.\quad
\end{eqnarray}
In the manner of paper \cite{Emelyanenko:2022}, let us consider
the one-constant approximation $K_{11}=K_{33}\equiv K$ for
simplicity. An equilibrium director ${\bf n}({\bf r})$
distribution should be obtained by independent minimization of
free-energy density (4), while an equilibrium polarization $P({\bf
r})$ distribution has already been obtained by independent
minimization of free-energy density (5) and is presented by
Eq.~(11).
\begin{figure}[h!]
\includegraphics[width=0.45\linewidth,clip]{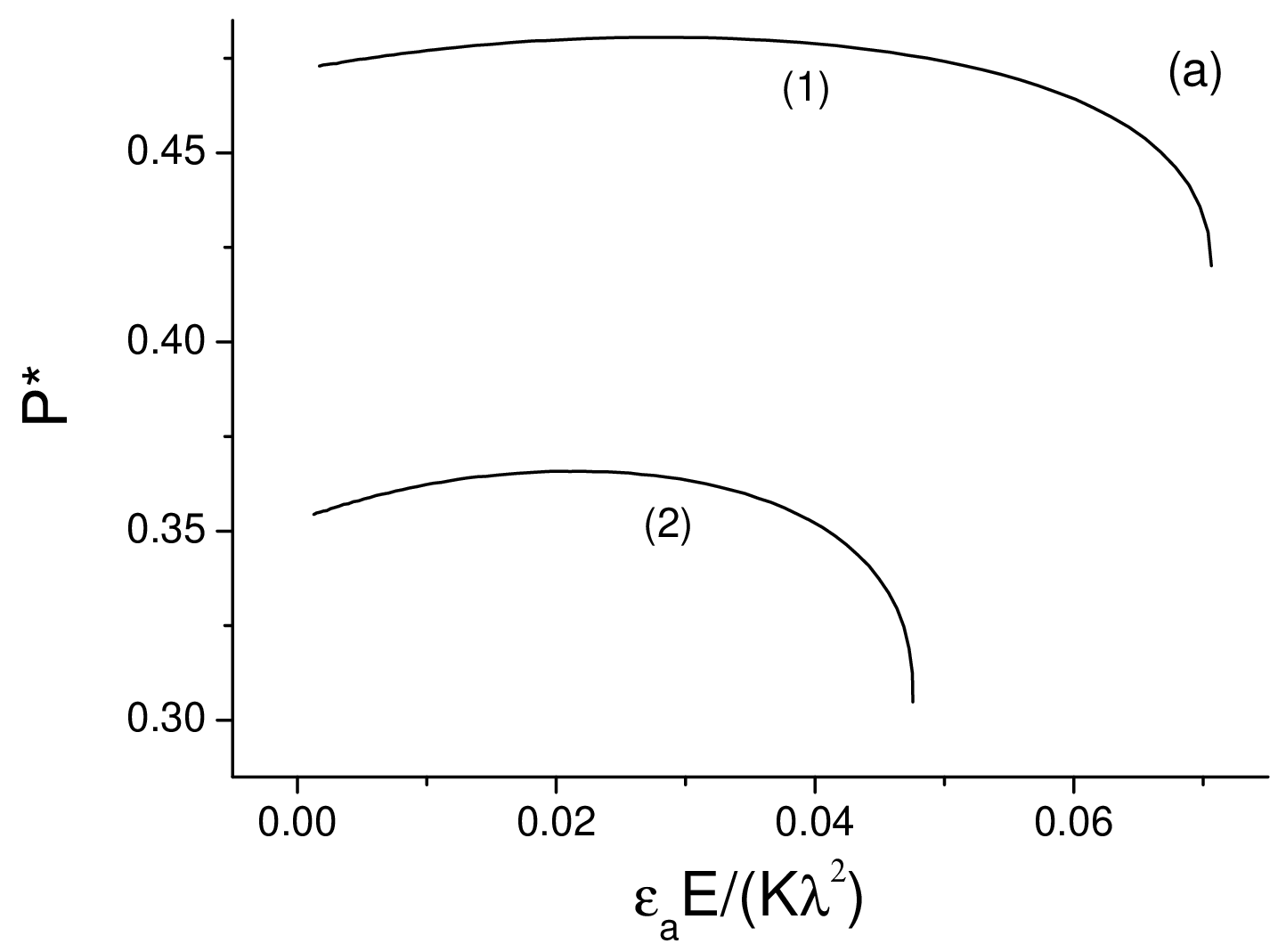}
\includegraphics[width=0.45\linewidth,clip]{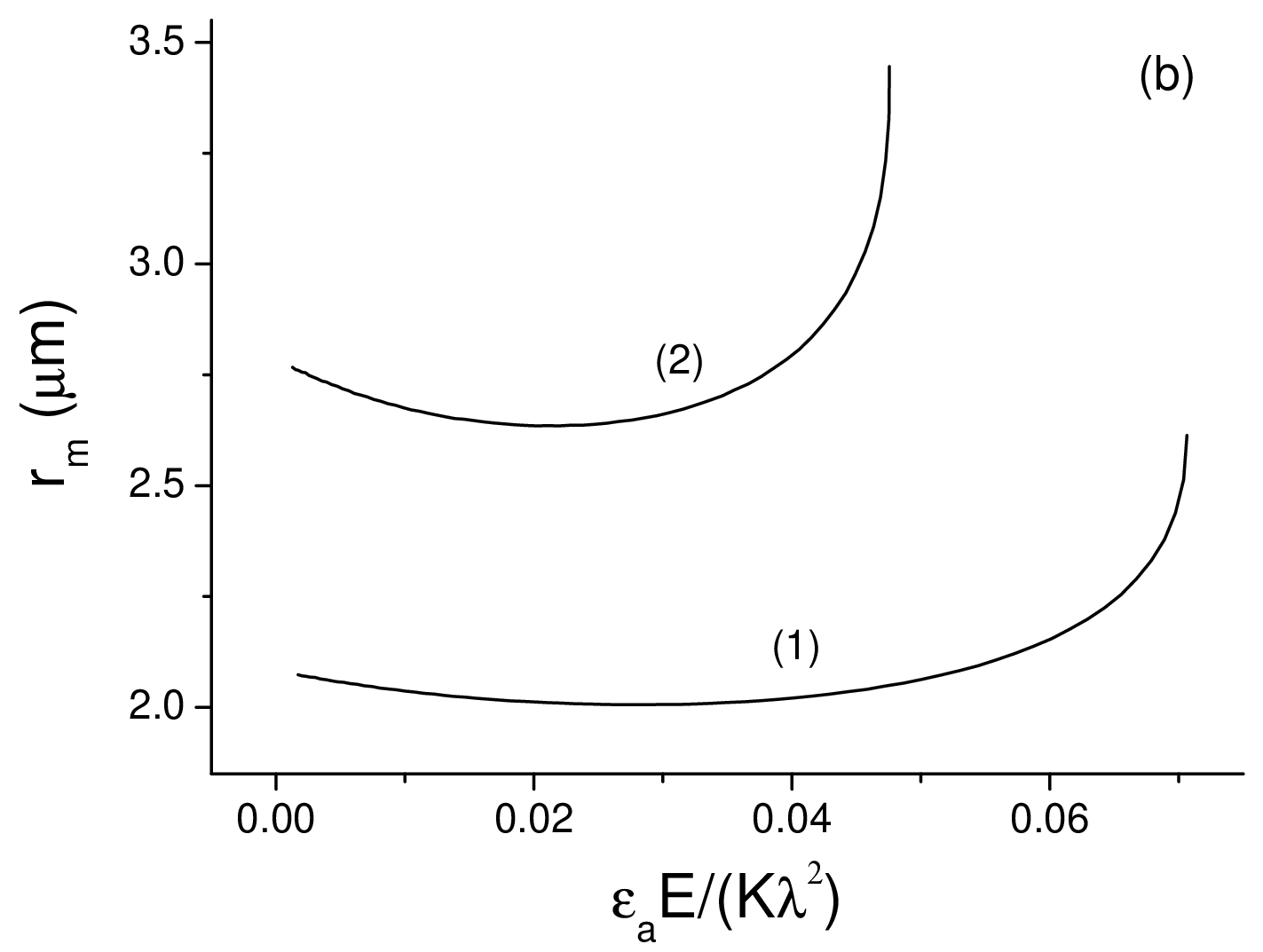}
\caption{\label{fig:epsart} Electric field dependencies of the
characteristic polar order parameter (a) and domain radius (b) in
$N_F^{1D}$ at $\sigma_0 J_{202}^{(0)}/k_B=2032\,K$, $\sigma_0
J_{101}^{(0)}/k_B=362\,K$, $\lambda=2\,\mu m^{-1}$,
$J_A/k_B=113\,K \mu m$, $K\,V_0=5\times 10^{-35}N\,m^3$ and
$T=70^{\circ}C$ [curves (1)] and $81^{\circ}C$ [curves (2)].}
\end{figure}
\begin{figure*}[tbh!]
\includegraphics[width=0.4\linewidth,clip]{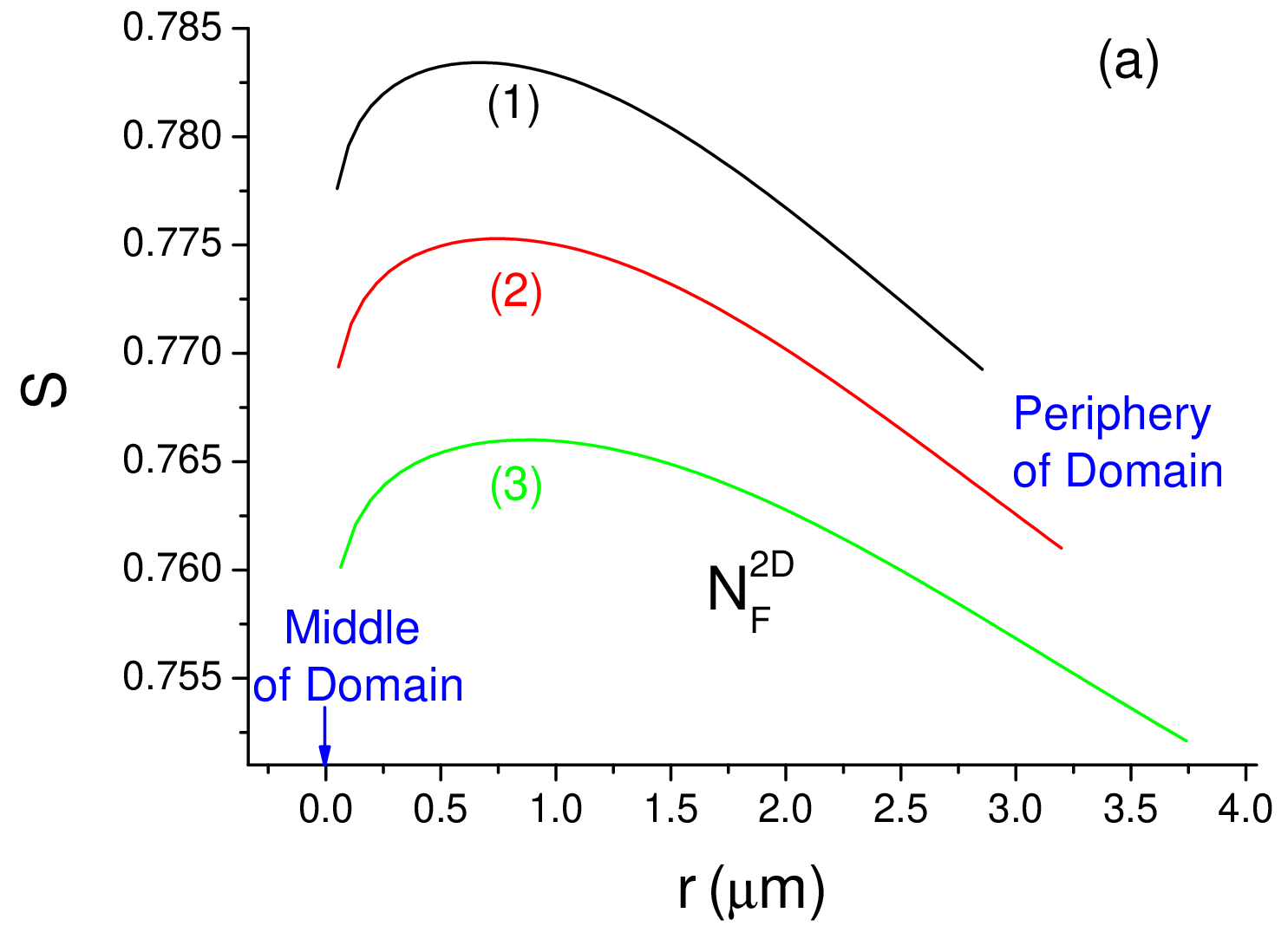}
\includegraphics[width=0.4\linewidth,clip]{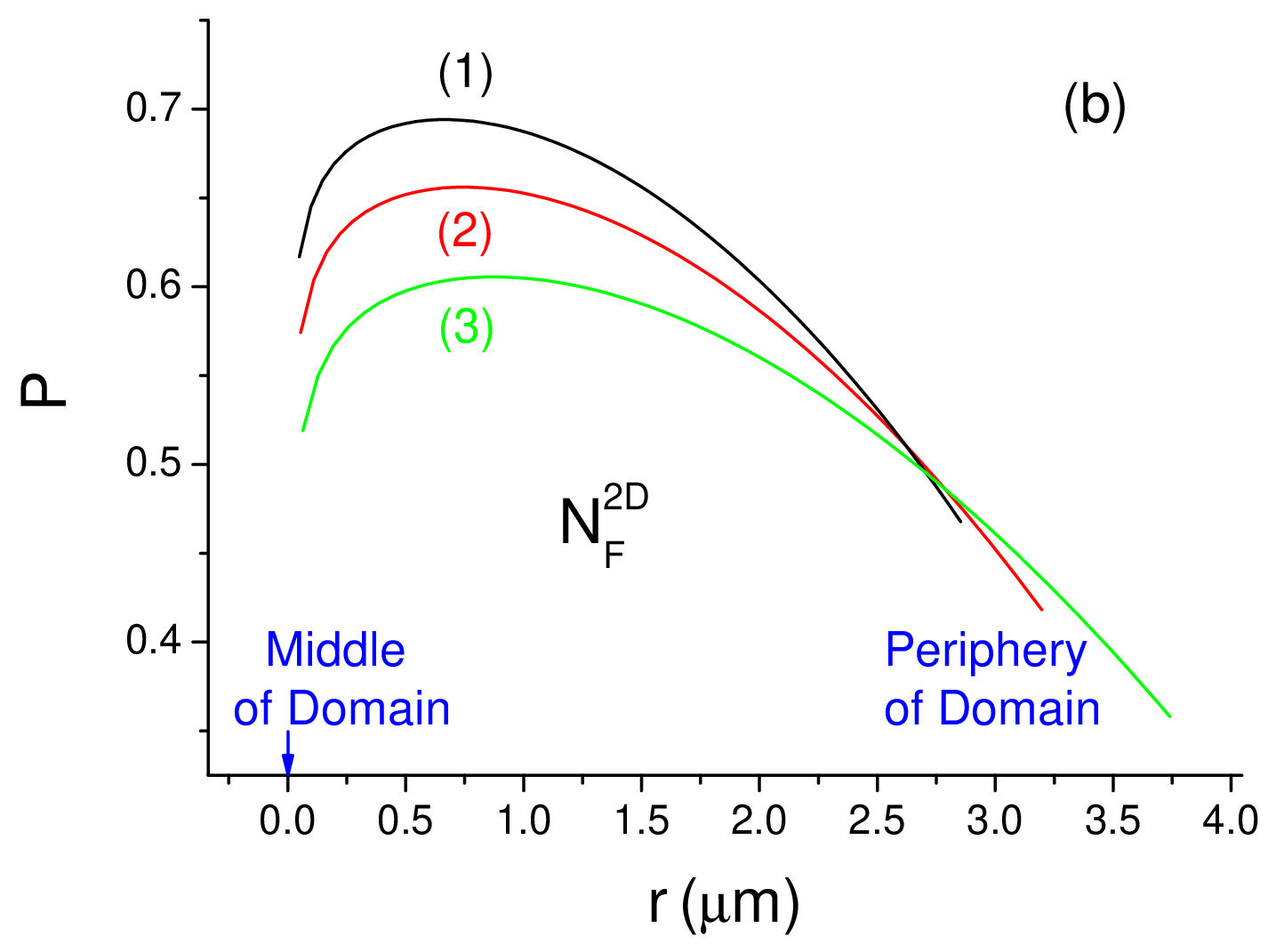}
\includegraphics[width=0.4\linewidth,clip]{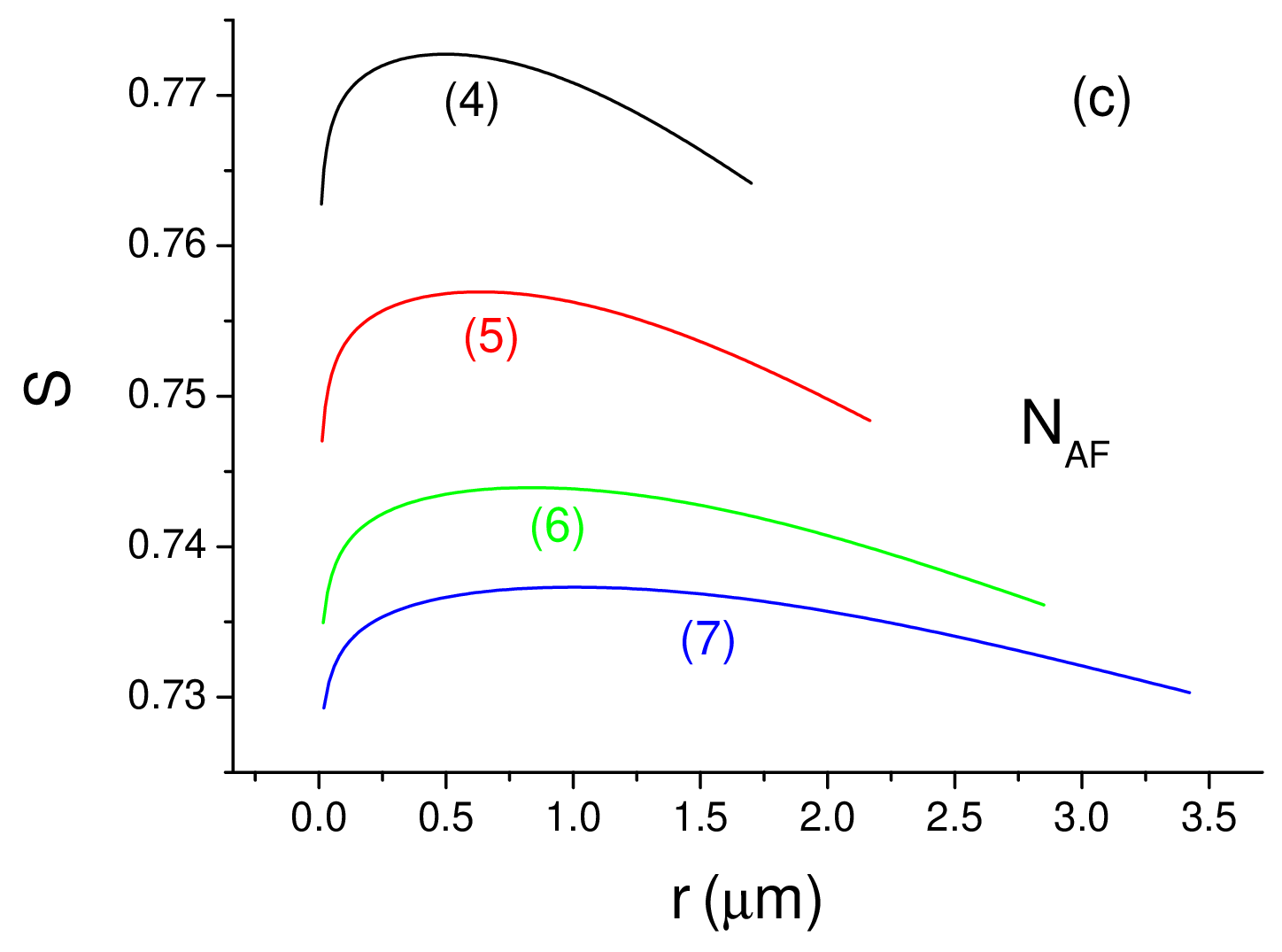}
\includegraphics[width=0.4\linewidth,clip]{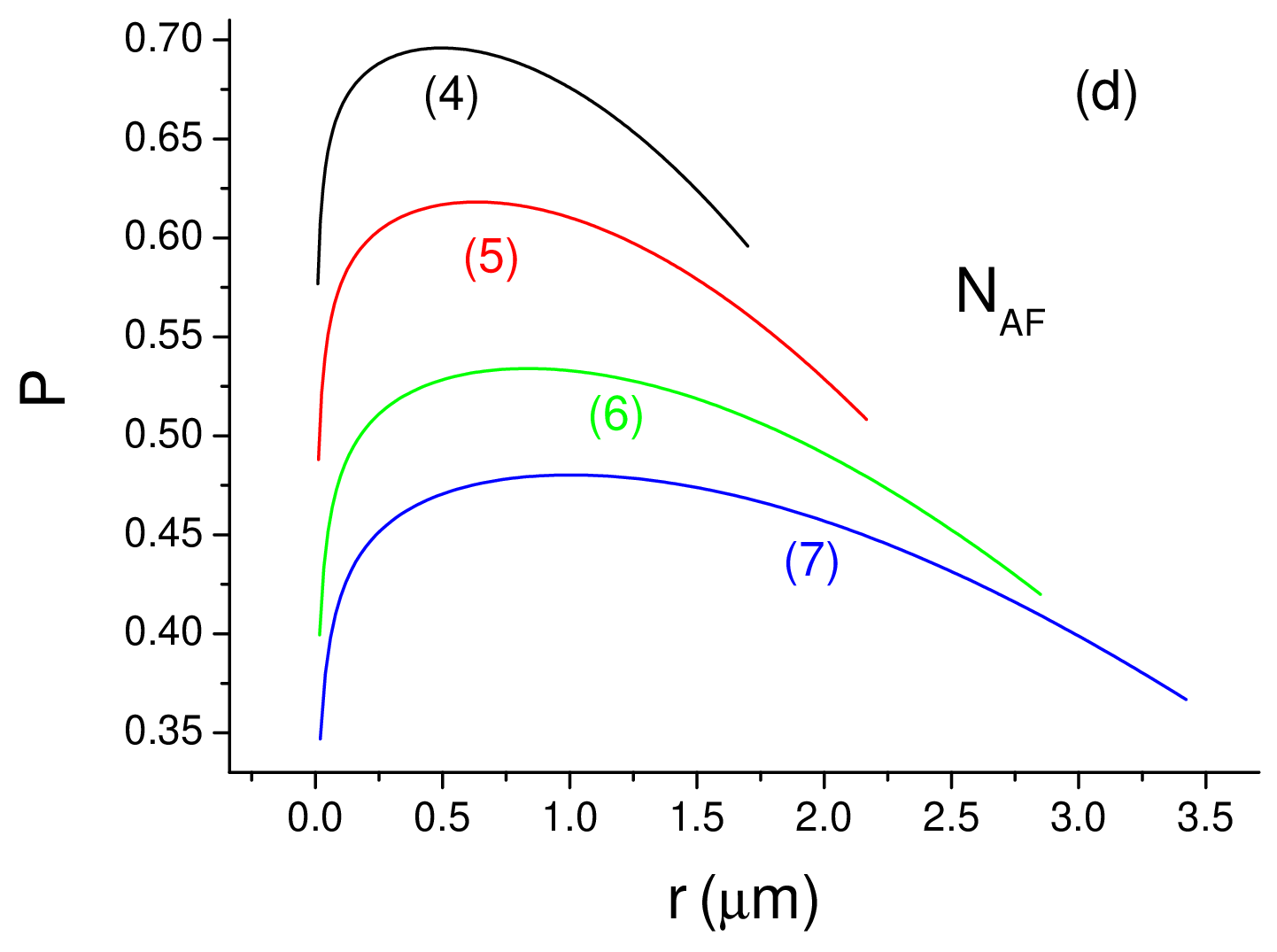}
\includegraphics[width=0.4\linewidth,clip]{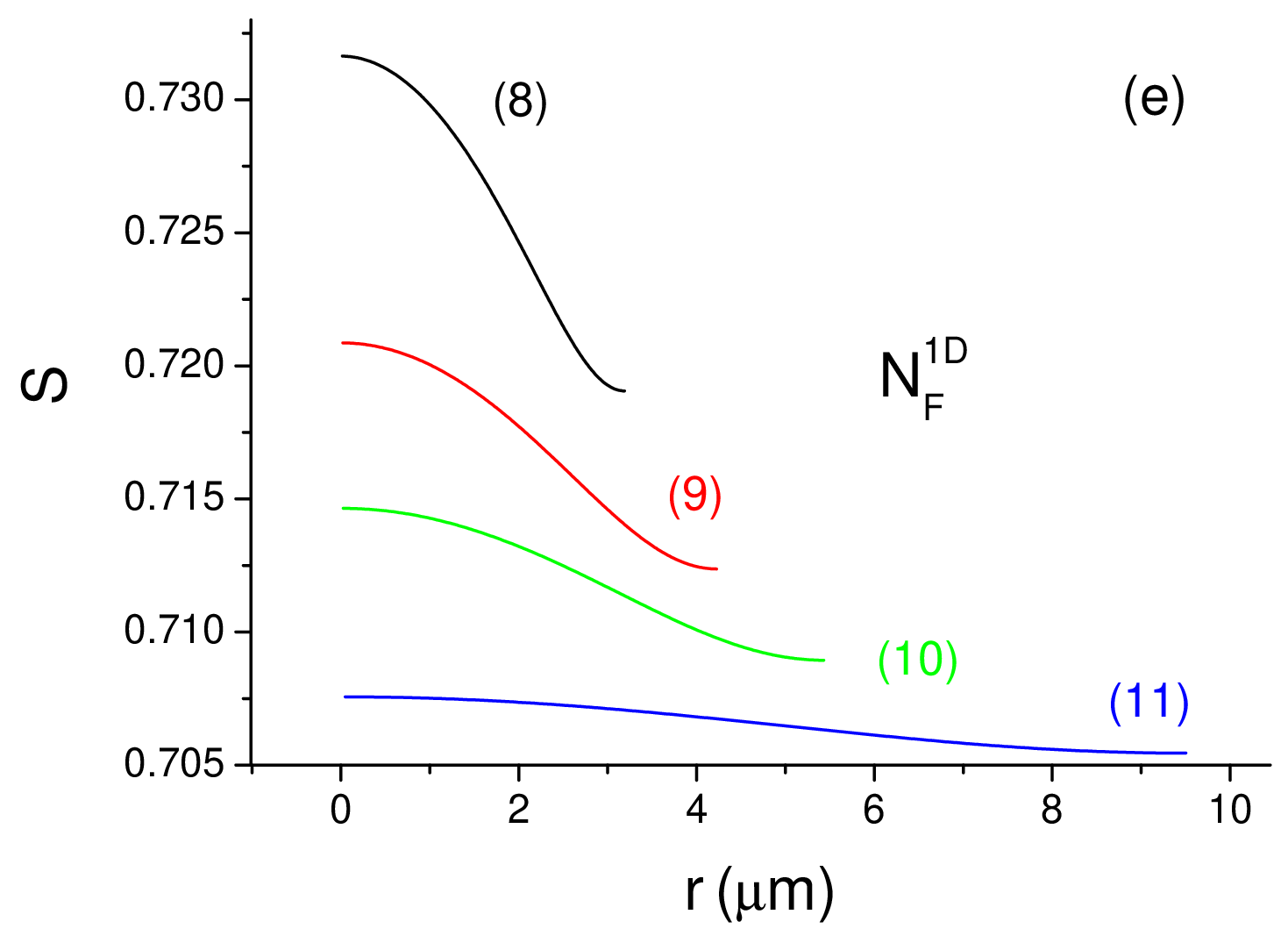}
\includegraphics[width=0.4\linewidth,clip]{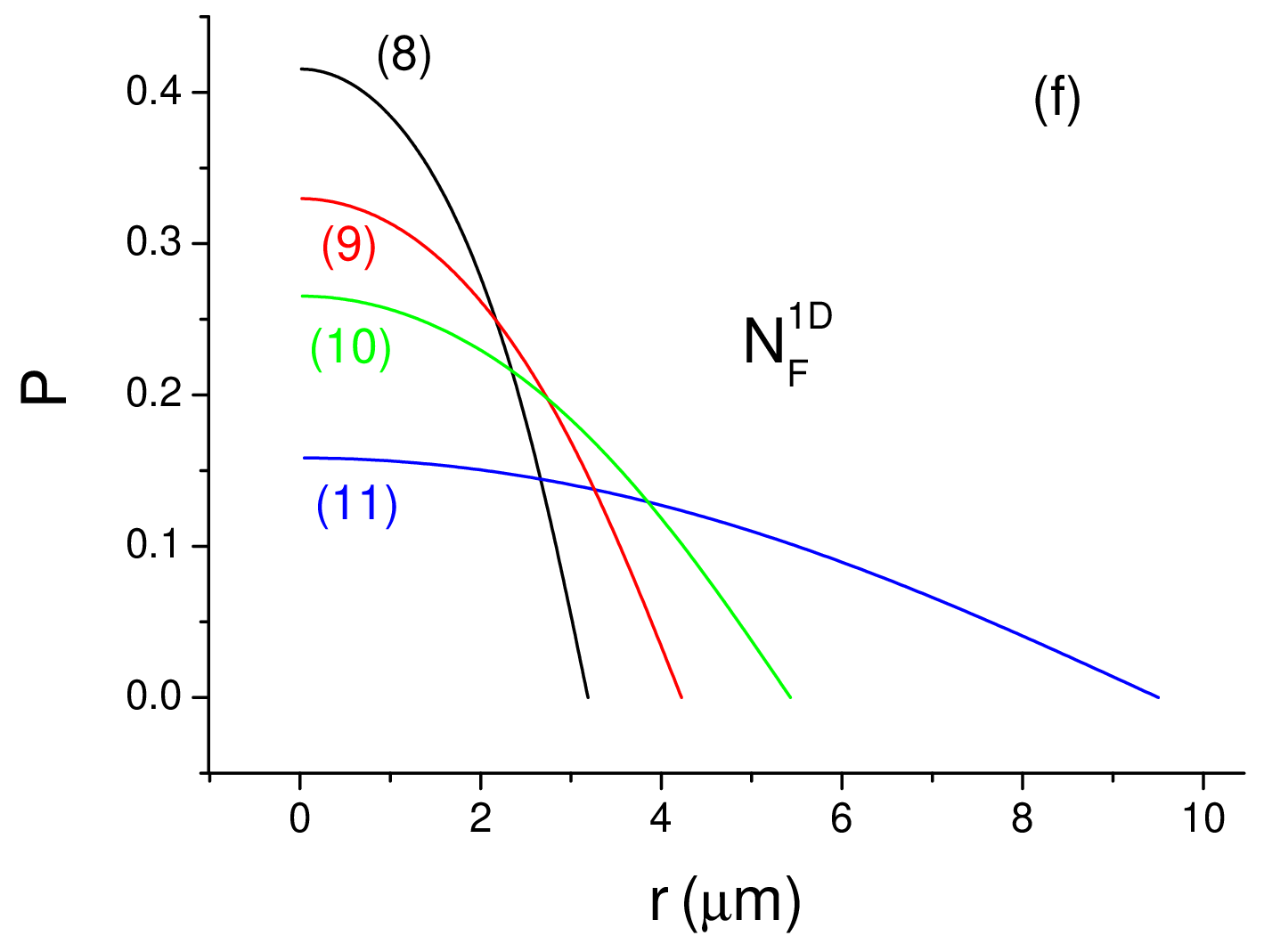}
\caption{\label{fig:epsart} Distribution of the $S$ [(a),(c),(e)]
and $P$ [(b),(d),(f)] orientational order parameters within the
domains in $N_F^{2D}$ [(a),(b)], $N_{AF}$ [(c),(d)] and $N_F^{1D}$
[(e),(f)] at $T=57^{\circ}C$ (1); $62^{\circ}C$ (2); $67^{\circ}C$
(3); $69^{\circ}C$ (4); $76^{\circ}C$ (5); $81^{\circ}C$ (6);
$83^{\circ}C$ (7); $85^{\circ}C$ (8); $88^{\circ}C$ (9);
$90^{\circ}C$ (10); $92^{\circ}C$ (11) at $E=0$, $\sigma_0
J_{202}^{(0)}/k_B=2032\,K$, $\sigma_0 J_{101}^{(0)}/k_B=362\,K$,
$\lambda=2\,\mu m^{-1}$, $J_A/k_B=113\,K \mu m$ and
$K\,V_0=5\times 10^{-35}N\,m^3$. Radius $r$ is defined in Fig.~2
for all the polar phases.}
\end{figure*}
Precise minimization of Eq.~(4) with constraint (11) appears to be
complicated. The complexity is in the fact that minimization with respect to director ${\bf n}$ and polarization value $P$ should be done independently, while varying in the space polarization $P$ (and this variation is unknown before we know the distribution of ${\bf n}$) participates in differential Eq.~(4), from where this distribution of ${\bf n}$ is supposed to be obtained. For this purpose, let us consider a perturbation theory based on the assumption that variation of the order parameters in space is small. In the framework of perturbation theory, let us first
consider the uniform polarization $P({\bf r})=P^*$ in Eq.~(4). In
both cases of ferroelectric domains presented in Figs.~3~(b) and
(c), let us consider variation of angle $\theta$ from zero (at the
${\bf x}$-axis of one domain) to $\pi$ (at the ${\bf x}$-axis of the
neighboring domain). Then the same simplification for the
free-energy density is valid for all the structures presented in
Fig.~2:  all the terms proportional either to $d\theta/d r$ or to
$\lambda$ should have opposite signs in the neighboring domains,
and the corresponding terms vanish in average. Taking this into
account, substituting Eq.~(15) into Eq.~(4), and minimizing
free-energy density (4) with respect to $\theta$ and $d\theta/dr$,
as presented, for example, in Ref.~\cite{Emelyanenko:2010},
Appendix A, one obtains the following equation of state:
\begin{equation}
\left(\frac{d \theta}{d
r}\right)^2+\tau^2|\cos\theta|-\frac{\delta}{r^2}\sin^2\theta=\frac{\tau^2}{k^2}
\quad,\quad
\end{equation}
where $\tau\equiv\sqrt{2\varepsilon_a E P^*/K}$ is the reduced
electric field, and $k$ is some constant independent of angle
$\theta$, which should be obtained by independent minimization of
the free-energy density. Introducing new dimensionless variable
$\psi\equiv \tau\, r/k$, one obtains from Eq.~(16):
\begin{equation}
\frac{d\theta}{d\psi}=\sqrt{1-k^2|\cos\theta|+\delta/\psi^2\,\sin^2\theta}
\quad.\quad
\end{equation}
Radius $r_m$ of the domain can now be found by minimization of the
free energy with respect to parameter $k$. This, however, can be
done in a more precise way partially taking into account the
non-uniformity of $P({\bf r})$. Indeed, from Eq.~(11) it
approximately follows [after expansion of the exponent in Eq.~(12)
in Taylor series with substitution of Eq.~(10)] that $P({\bf
r})\sim({\mbox {\boldmath $\nabla$}}\cdot{\bf n})+\varepsilon_a
E\cos\theta/(K\lambda)$, where the first term is flexoelectric
polarization and the second term is induced by electric field
polarization. One notes from Eq.~(4) that any re-scale of
coordinate $r$, at which $r\lambda P$ and $\tau/(\lambda P)$
remain constant, does not change the free-energy density. This
means in the end that distribution of polarization is determined
only by distribution of angle $\theta$ in the space. Let us
therefore write the following trial approximation for the polar
order parameter:
\begin{eqnarray}
P(\theta)=P^* r_m\{({\mbox {\boldmath $\nabla$}}\cdot{\bf
n})+\tau^2\cos\theta/(2\lambda P^*)\}  =P^*
\psi_m\biggl\{\frac{\delta}{\psi}\sin\theta+\cos\theta\biggl(\frac{d\theta}{d\psi}+\frac{1}{2}k\tilde{\tau}\biggr)\biggr\}
\quad.\quad
\end{eqnarray}
\begin{figure}[h!]
\includegraphics[width=0.45\linewidth,clip]{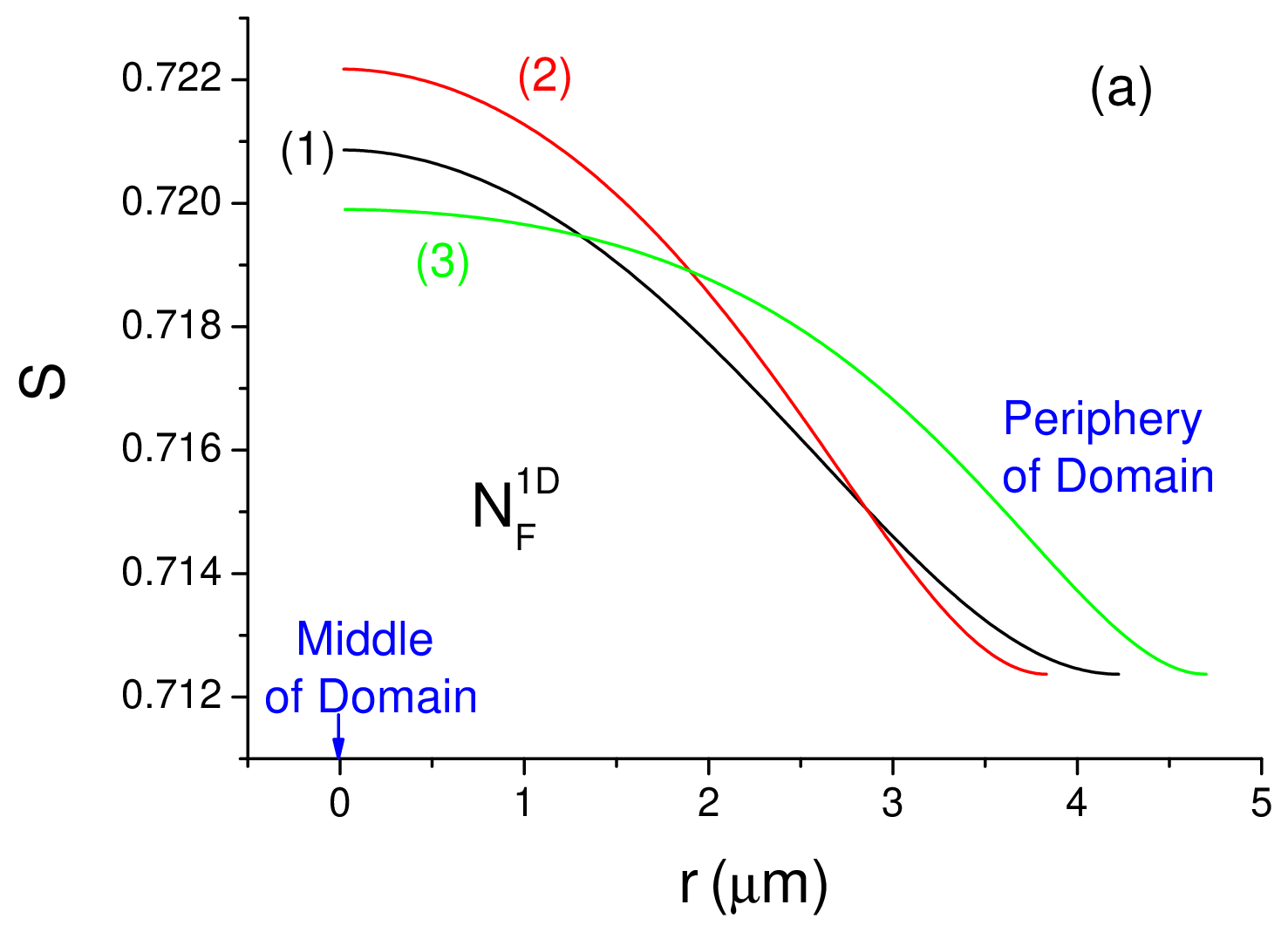}
\includegraphics[width=0.45\linewidth,clip]{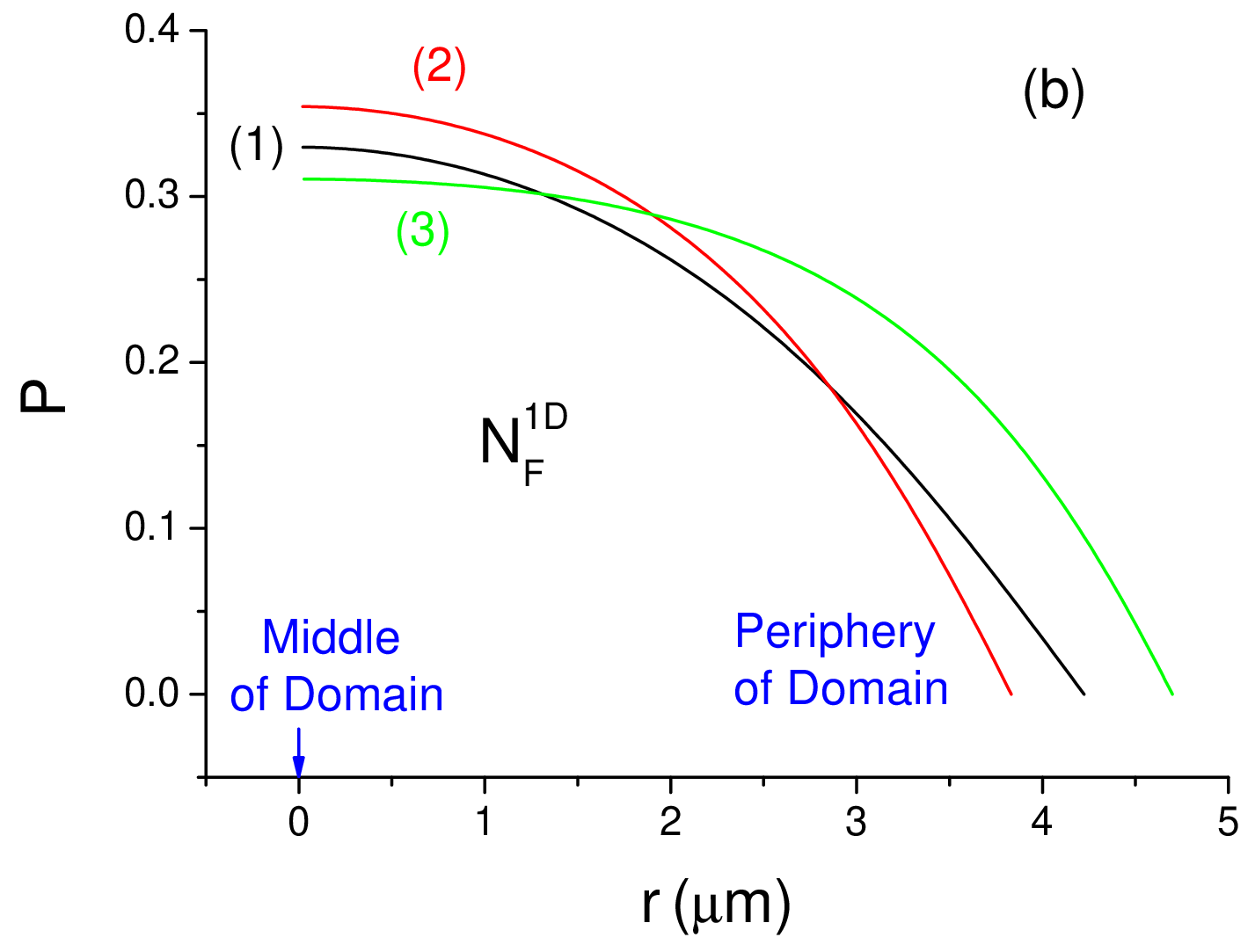}
\caption{\label{fig:epsart} Distribution of the $S$ (a) and $P$
(b) orientational order parameters within the domains in
$N_F^{1D}$ at $\varepsilon_a E/(K\lambda^2)=0$ (1); $0.01$ (2);
$0.028$ (3) at $T=87^{\circ}C$, $\sigma_0
J_{202}^{(0)}/k_B=2032\,K$, $\sigma_0 J_{101}^{(0)}/k_B=362\,K$,
$\lambda=2\,\mu m^{-1}$, $J_A/k_B=113\,K \mu m$ and
$K\,V_0=5\times 10^{-35}N\,m^3$. Radius $r$ is defined in
Fig.~2~(c).}
\end{figure}
One notes from Eq.~(18) that, in the case of $N_F^{2D}$, the $P^*$
proportionality coefficient coincides with $P$ at $\theta=\pi/2$
(polar order parameter at periphery $r_m$ of ferroelectric
domain). In the cases of $N_F^{2D}$ and $N_{AF}$, the $P^*$
coefficient formally corresponds to different place $r^*$ within
the domain, other than periphery $r_m$. Regardless of the kind of
the domain, however, the expression in figure brackets in Eq.~(18) is equal to $1/r_m$ at $r^*$.
Substituting Eqs.~(15), (17) and (18) into Eq.~(4), integrating
the free-energy density along the radius of domain (with the
$r\,dr$ Jacobian for 2D-splay or with the $dr$ Jacobian for
1D-splay) and dividing the result by the cross section area (for
2D-splay) or by the length of domain (for 1D-splay), one obtains
the expression for the average free-energy density, which should
be farther minimized with parameter $k$. This could be done for
each polar nematic phase similarly to that presented in
Ref.~\cite{Emelyanenko:2022} for $N_{AF}$. Subsequently, at any
value of $\tilde{\tau}$, the $r(\theta)$ dependence can be
obtained, and, in particular, radius $r_m$ of the domain can be
obtained. In $N_{AF}$, $r_m\lambda P^*\approx 1.55$ and maximum
tilt is $\theta_m\approx 64^\circ$. In $N_F^{1D}$ and $N_F^{2D}$,
radius $r_m$ of the domain generally depends on the applied
electric field, while maximum tilt is always equal to
$\theta=\pi/2$. Several $\theta(r)$ and $({\mbox {\boldmath
$\nabla$}}\cdot{\bf n})$ dependencies at several particular values
of dimensionless electric field $\tilde{\tau}$ are presented in
Figs.~10~(a) and (b), respectively, for $N_F^{2D}$ [blue curves
(1) and (2)] and $N_F^{1D}$ [red curves (3), (4) and (5)]. One
notes that the tilt of director varies almost linearly in both
$N_F^{2D}$ and $N_F^{1D}$, with a slight tendency to greater
variation in the middle of each domain in $N_F^{2D}$ and,
oppositely, at the domain periphery ($r=r_m$) in $N_F^{1D}$. From
Fig.~10 it follows that, at moderate values of electric field, the
maximum splay deformation in $N_F^{1D}$ is achieved at $\theta=0$,
at which the director is parallel to electric field. This is the
configuration, at which both flexoelectric and induced
polarizations give optimal summarized contribution to the free
energy. Therefore both $N_F^{2D}$ and $N_{AF}$ exhibit a
transition into $N_F^{1D}$ at application of electric field.
However, there always exists a disbalance between the induced and
flexoelectric polarizations. Indeed, the flexopolarization can
exist only in the presence of director deformation. However, at
application of electric field, the structure becomes more uniform,
and the splay deformation reduces. Therefore, at higher electric
field, the maximum in Fig.~10~(b) reduces and shifts to the
position, where director is not parallel to electric field. At
$\tilde{\tau}\approx 0.843$, the splay phase becomes unstable, and
a transition into paraelectric nematic phase happens. The electric
field dependencies of characteristic polar order parameter $P^*$
and domain radius $r_m$ at particular fixed temperatures within
$N_F^{1D}$ are presented in Figs.~11~(a) and (b), respectively.
Both dependencies are generally not monotonic because of the
nontrivial correlation between splay and electric field. At higher
value of electric field, $P^*$ greatly decreases and $r_m$ greatly
increases just before the transition into $N$.

Knowing the director distribution in space in $N_F^{2D}$, $N_{AF}$
and $N_F^{1D}$, one obtains the distributions of $S$ and $P$ order
parameters in each phase using Eqs.~(10)--(12). For this purpose,
one should substitute approximation (18) into
Eq.~(10). In particular, at specific places $r^*$ within each
domain, where polar order parameter $P$ coincides with coefficient
$P^*$, one immediately obtains that the whole expression in figure
brackets in Eq.~(10) is equal to
$\left[J_{110}^{(1)}+J_{011}^{(1)}\right]/(6r_m)$. From recurrent
Eqs.~(11)--(12) one obtains $S^*$ and $P^*$ first, and then the
whole distribution of $S(r)$ and $P(r)$ within the domain, which
are presented in Fig.~12. One notes that both $S$ and $P$
generally decrease with the increasing temperature. The maximal
values of both parameters are observed close to the middle of
domain in $N_F^{2D}$ and $N_{AF}$ and in the middle exactly in
$N_F^{1D}$. Polar order parameter reaches zero at the periphery of
each domain in $N_F^{1D}$, while in $N_F^{2D}$ and $N_{AF}$ it
does not, which means that flexopolarization exhibits step-wise
reversal between the domains without director disruption.
Distributions of $S(r)$ and $P(r)$ at several non-zero values of
electric field are also presented in Fig.~13. One notes that
profiles of both $S$ and $P$ first tend to become sharper at
moderate electric field and then smoother at higher electric
field.

\subsection{Computer simulations}
\begin{figure}[h!]
\includegraphics[width=0.45\linewidth,clip]{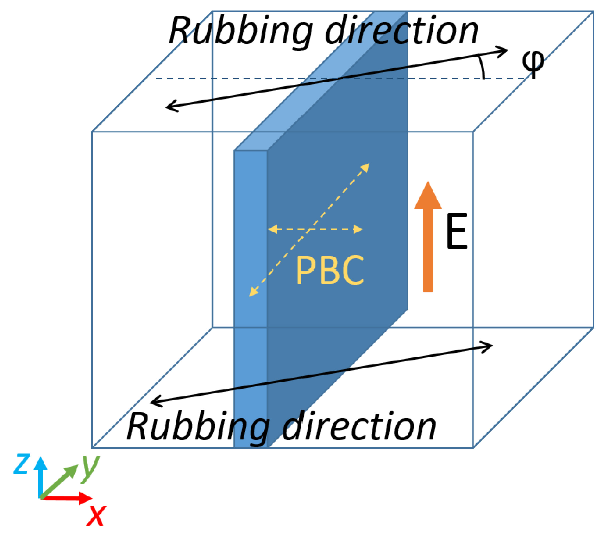}
\caption{\label{fig:sim-geometry} Principal geometry of the
simulations box and the polar nematic film. Black arrows show the
rubbing direction of planar alignment of the film. Orange
arrow shows the electric field direction. Yellow dash arrows show
the periodic boundary condition directions of the simulation box.}
\end{figure}
\begin{figure}[h!]
\includegraphics[width=0.45\linewidth,clip]{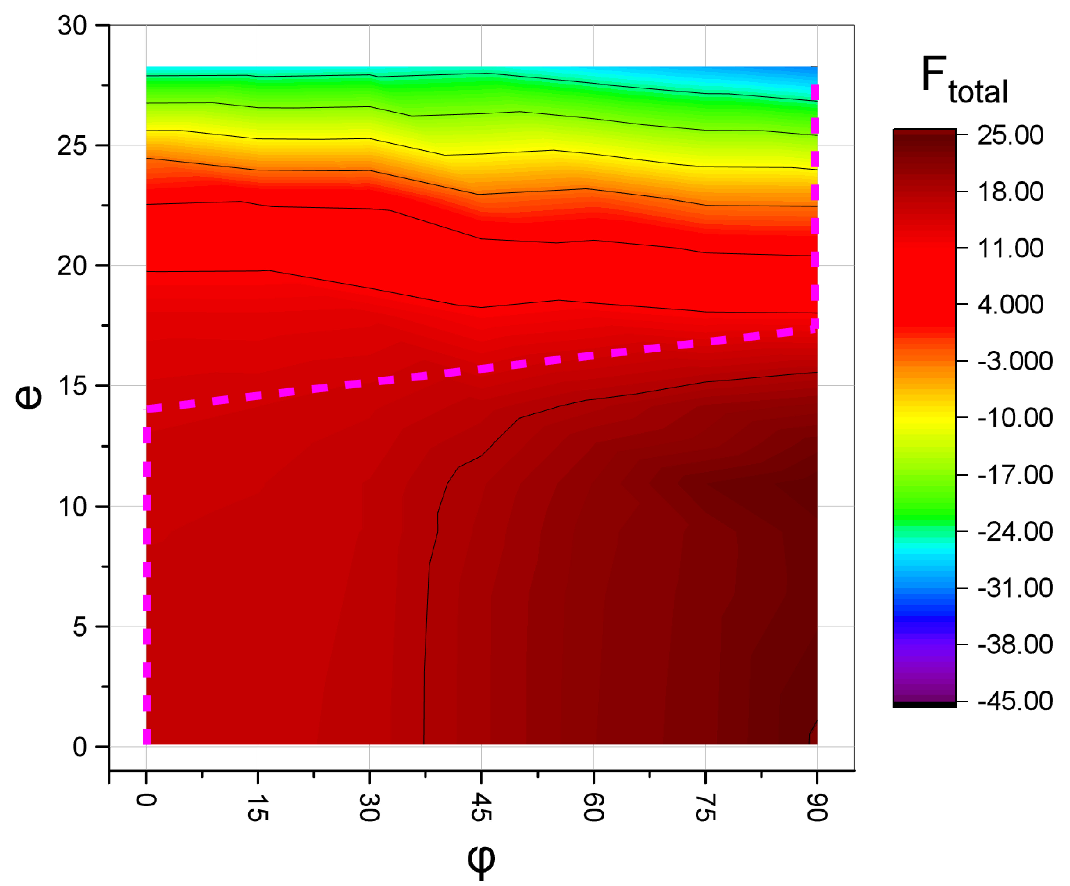}
\caption{\label{fig:sim-energies} Dependence of the total free
energy on the value of dimensionless electric field $e$ and
rubbing direction orientation $\varphi$. Violet dash line traces
energy minimum over $e$.}
\end{figure}
\begin{figure*}[tbh!]
\includegraphics[width=0.95\linewidth,clip]{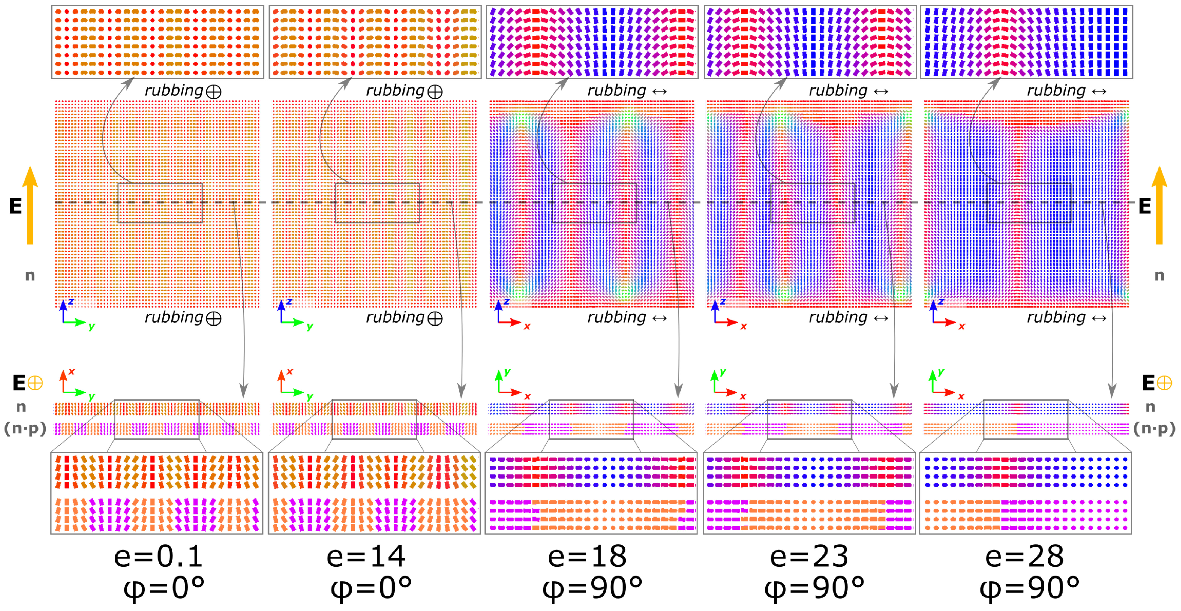}
\caption{\label{fig:sim-structures} Director $\mathbf{n}$ and
polarizability $(\mathbf{n}\cdot\mathbf{p})$ distributions at
various dimensionless electric field $e$ values in the central
cross-cuts perpendicular to the film plane (top) and along the
film plane (bottom). Director distributions are shown in color,
corresponding to the direction of $\mathbf{n}$ ($x$ - red, $y$ -
green, $z$ - blue). Polarizability is shown in
orange [$(\mathbf{n}\cdot\mathbf{p})=1$] and violet
[$(\mathbf{n}\cdot\mathbf{p})=-1$]. }
\end{figure*}
To perform calculations of director distribution in a polar
nematic film under the action of electric field, we have modified the
existing extended Frank elastic continuum approach
\cite{Emelyanenko:2022}, previously used for calculations of
polar nematic material.
The original approach takes into account the effects of director field
distortion with the $\lambda(\mathbf{n}\cdot\mathbf{p})$ term
included, as well as the formation of defects and finite energy of the
surface boundaries. In this paper, we have modified the free energy to
take into account the action of an electric field:
\begin{eqnarray}
F=\frac{1}{2}\int\limits_V\biggl\{K_{11}({\bf n}({\mbox {\boldmath
$\nabla$}}\cdot{\bf n})-\lambda\,{\bf p})^2
+K_{22}({\bf n}\cdot[{\mbox {\boldmath $\nabla$}}\times{\bf n}])^2
+K_{33}[{\bf n}\times[{\mbox {\boldmath $\nabla$}}\times{\bf
n}]]^2 \biggr\} dV \nonumber\\
-\varepsilon_a\int\limits_V(\mathbf{n}\cdot\mathbf{p})(\mathbf{n}\cdot\mathbf{E})
dV
 + \frac{W}{2}\int\limits_\Omega (1-\cos^2\gamma)
d\Omega +F_{\rm def} \quad,\quad
\end{eqnarray}
where $K_{11}$, $K_{22}$ and $K_{33}$ are the splay, twist and
bend elastic constants, respectively, $K_{11}\lambda$ is the
flexoelectric constant, $\mathbf{p}$ is polarizability direction
vector, $\mathbf{E}$ is electric field intensity, $V$ is the bulk
of the sample having surface $\Omega$, $W$ is the surface
anchoring energy density, $\gamma$ is the angle between local
director and normal to the surface, $F_{\rm def}$ is the energy of
defects calculated by the summation of the point and linear defect
energies (see the details in Ref. \cite{Rudyak:2013}). The details
of optimization are presented in Ref.~\cite{Emelyanenko:2022}. For
simplicity, polarizability direction vector $\mathbf{p}$ is
supposed to be a unit vector parallel or anti-parallel to
$\mathbf{n}$ in each point. In addition, in correspondence with
theoretical part of the paper, an algorithm accepted only those
steps with
$(\mathbf{n}\cdot\mathbf{p})(\mathbf{n}\cdot\mathbf{E})\geq0$. As
a result, our simulation annealing procedure leads to minimization of
the free energy over both director $\mathbf{n}$ and polarizability
direction $\mathbf{p}$ distributions in a self-consistent way.

The one-constant approximation was used for simplicity:
$K_{11}:K_{22}:K_{33} = 1:1:1$, and the value of $\lambda$ is set
to 10. To take into account the potential formation of
disclination lines, their cores linear energy density was set to
$f_{core}^{line} = 10 K_{11}$. The cubic simulation box of size
$0.125\times2\times2$ was rendered into $4\times64\times64$
lattice. For $x$ and $y$ facets, the periodic boundary conditions
were applied. For $z$ facet, planar aligned boundary conditions
were set with rubbing direction having angle
$\varphi\in[0^\circ;90^\circ]$ with the $x$ axis and $\mu_1 =
Wd/K_{11} = 400$, where $d$ is the film thickness (see
Fig.\,\ref{fig:sim-geometry}).
The electric field $\mathbf{E}$ was oriented perpendicular to the
film plane, and the value of dimensionless electric field
intensity $e=Ed(\frac{\varepsilon_a}{K_{11}})^{1/2}$ varied from
0.1 to 30. For each $e$, we produced $6.1\times10^{10}$ steps
($3\times10^7$ parallel multisteps) Monte-Carlo annealing
optimization with 4 independent runs to find the energy-optimal
structures.

The resulting structures strongly depend on the value of electric
field and the rubbing direction $\varphi$.
Fig.\,\ref{fig:sim-energies} shows the dependency of the total
free energy of the system on these two parameters. At low value of
electric field ($e$ from 0.1 to 14), the energy-optimal structure
corresponds to the antiferroelectric splay in the $xy$ plane
(supposed to be the plane of the substrate in real experiment)
with alternating sign of polarizability. The average director
orientation is almost parallel to the rubbing direction. Some
slight ferroelectric modulation is also present: a projection of
director along electric field arises in the middle of each
antiferroelectric domain, so that the projection of director lines
on the $zy$ plane (perpendicular to rubbing) gains the shape of
periodical arcs. This structure is visualized as longitudinal to
the rubbing direction stripes (Fig.\,\ref{fig:sim-structures}).
Above the threshold value $e*\approx15$, the system undergoes a
transition related to the reorientation of the arcs along the
rubbing direction, and the arcs themselves become much bigger,
while the antiferroelectric modulation in the $xy$ plane
disappears. This structure (at $e>15$) is visualized as transverse
to the rubbing direction stripes (Fig.\,\ref{fig:sim-structures}).
At further increasing electric field, the layer period starts
growing [similarly to that in theory, see Fig.~11~(b)], and the director divergence in the middle each domain decreases [similarly to that in theory, see Fig.~10~(b)]. Computer simulations describe well the transformation
from antiferroelectric to ferroelectric structure with the
reorientation of stripes, which is observed experimentally and
presented in Sec. III~A.

\section{Conclusion}

The origin and structures of ferroelectric and antiferroelectric
splay nematic phases are outlined. The double-splay ferroelectric
$N_F^{2D}$ and antiferroelectric $N_{AF}$ nematic phases are
composed of quasi-cylindrical periodical domains. Without electric
field, $N_F^{2D}$ and $N_{AF}$ are observed in the
lower-temperature range. The single-splay ferroelectric $N_F^{1D}$
nematic phase is composed of planar periodical domains. Without
electric field, $N_F^{1D}$ is observed in the higher-temperature
range. In the presence of electric field, all the splay nematic
phases first (at moderate electric field) transform into
$N_F^{1D}$ and then (at higher electric field) -- into
paraelectric nematic phase $N$ having uniform director
orientation. The origin of all the splay nematic phases is
flexoelectric effect due to the polarity of molecules. The origin
of the transformations between phases in electric field is
the non-trivial interplay between flexoelectric and induced
polarizations. The distribution of director and both polar $P$ and
non-polar $S$ orientational order parameters within the domains of
all the splay nematic phases is found. Variation of the structure and
properties of the splay nematic phases with variation of
temperature and electric field are investigated. The electric
field -- temperature phase diagram is obtained. The equilibrium
domain size was found to increases and polarization was found to
decrease in each polar phase with the increasing temperature.
Several additional phase transitions related to optimization of
the domains within the cell gap were found and explained.

\acknowledgments A.V.E. and V.Yu.R. thank the Russian
Foundation for Basic Research (project No. 21-53-50008) for the
financial support of theoretical investigation presented in this
work. F.A, H.N and K.I. thank Japan Society for the Promotion of
Science (project No. JPJSBP120214814) for the financial support of
experimental investigation presented in this work. The research
was carried out using the equipment of the shared research
facilities of HPC computing resources at Lomonosov Moscow State
University. The authors are grateful to S.A. Shvetsov for help.

\end{document}